\documentclass[twocolumn,showpacs,preprintnumbers,amsmath,amssymb]{revtex4}

\usepackage{graphicx}
\usepackage{dcolumn}
\usepackage{bm}

\begin{document}

\title{Novel anisotropic spin singlet pairings in Cu$_x$Bi$_2$Se$_3$ and Bi$_2$Te$_3$}

\author{Lei Hao$^{1}$, Guiling Wang$^{1}$, Ting-Kuo Lee$^{2}$, Jun Wang$^{1}$, Wei-Feng Tsai$^{3}$, and Yong-Hong Yang$^1$}
 \address{$^1$Department of Physics, Southeast University, Nanjing 210096, China \notag \\
 $^2$Institute of Physics, Academia Sinica, Nankang, Taipei 11529, Taiwan \notag \\
 $^3$Department of Physics, National Sun Yat-sen University, Kaohsiung 804, Taiwan}

\date{\today}

\begin{abstract}
Possible anisotropic spin singlet pairings in Bi$_2$X$_3$ (X is Se or Te) are studied.
Among six pairings compatible with the crystal symmetry,
two novel pairings show nontrivial surface Andreev bound states, which form flat bands
and could produce zero bias conductance peak in measurements like point contact spectroscopy.
By considering purely repulsive short range Coulomb interaction
as the pairing mechanism, the dominant superexchange terms are all antiferromagnetic,
which would usually favor spin singlet pairing in Bi$_2$X$_3$.
Mean field analyses show that the interorbital pairing interaction favors a mixed spatial-parity anisotropic pairing state, and one pairing channel with zero energy surface states has a sizable component. The results provide important new information for future experiments.
\end{abstract}


\maketitle

\section{\label{sec:Introduction}Introduction}

Cu$_x$Bi$_2$Se$_3$ is the first superconductor emerging from a three dimensional topological insulator (TI).\cite{hor10,wray10} As a bran-new superconducting (SC) material with topologically nontrivial normal state, it has been suspected to be a time reversal invariant (TRI) topological superconductor (TSC), which supports Andreev bound states (ABS) on the surface.\cite{fu10,hao11,sasaki11,hsieh12,yamakage12} The surface ABS in TSCs are massless Majorana fermions, which are novel particles identical to their antiparticles and are under intensive search also because of their prospect of application in topological quantum computing.\cite{wilczeketal}

Based on a phenomenological on-site attractive interaction, Fu and Berg indeed find some triplet pairings that support surface ABS in Cu$_x$Bi$_2$Se$_3$.\cite{fu10} Subsequent point contact spectroscopy experiments observed zero bias conductance peaks (ZBCPs), providing smoking-gun evidence for the existence of surface ABS.\cite{sasaki11} Though recently there are reports advocating conventional $s$ wave pairing\cite{levy12}, the majority of experiments are indicating the unconventional nature of the pairing in Cu$_x$Bi$_2$Se$_3$.\cite{sasaki11,koren11,kirzhner12,chen12,kriener11,das11,bay12} For examples, a specific heat measurement on one hand suggests a fully gapped pairing while on the other hand is not in full agreement with BCS prediction.\cite{kriener11} Meissner effect measurements show an unusual field dependence of the magnetization, which is argued to be consistent with odd-parity spin triplet pairing.\cite{das11} Upper critical field measurements show the absence of Pauli limiting effect, contrary to conventional isotropic s wave pairings, pointing to a very likely triplet pairing.\cite{bay12}

There are definitely more works needed to identify the genuine pairing symmetries of Cu$_x$Bi$_2$Se$_3$ and the pressure induced SC state of Bi$_{2}$Te$_{3}$ and Bi$_{2}$Se$_{3}$.\cite{zhang11pa,zhang11pb,kirshenbaum13} One important question, which is by and large disregarded up to now, is the possible relevance of anisotropic spin singlet pairings to these new superconductors.\cite{hao11,bay12} Since the anisotropic spin singlet parings are more commonly realized than triplet pairings among all unconventional superconductors, including the well-known cuprates and iron pnictides, it is highly desirable to explore this possibility. In this direction, one interesting open question is whether a possible anisotropic spin singlet paring can support surface ABS that could give a ZBCP in point contact spectra and scanning tunneling microscopy (STM) experiments.  Since the observation of ZBCP is considered to be a clear indication of possible unconventional pairing, it is very important to identify theoretically this possibility.

The above topic is interesting also in the symmetry classification of superconductors. Since spin-orbit interaction is important for Bi$_2$X$_3$ (X is Se or Te), TRI pairings all belong to the DIII symmetry class.\cite{schnyder08} In this symmetry class, while known topologically nontrivial pairings are all spin triplet\cite{schnyder08,kitaev09,sato09,fu10}, it is unclear if spin singlet pairings can also have nontrivial topological properties. The present system of Bi$_2$X$_3$ provides a good candidate to explore this possibility. In the work presented below we show that, when anisotropic pairing is considered, two interorbital singlet pairings give zero energy surface states and hence ZBCP in point contact spectra and STM experiments.

The remaining part of the paper is organized as follows. In Sec. II we construct a tight binding model by using group theory and mapping to an existing minimal model defined close to the Brillouin zone (BZ) center. Then we analyze the possible anisotropic spin singlet pairings from purely symmetry considerations, with an antiferromagnetic exchange term in the mind as a phenomenological mechanism for giving these pairings. The gap structures and spectral properties of the various pairings are then analyzed in detail. In Sec. III, we start from a purely repulsive short range Coulomb interaction and derive the dominant superexchange channels relevant to Bi$_{2}$Se$_{3}$ and Bi$_{2}$Te$_{3}$. Then a self-consistent mean field calculation in terms of a $t$$-$$U$$-$$V$$-$$J$ model is performed, which shows that the actual pairing is always a mixture of several components of the anisotropic singlet pairings identified in Sec. II. In Sec. IV, we discuss the implications of our work for experiments and then give a summary of the results. More technical details related to the results are provided in the Appendices.

\section{\label{sec:result1}singlet pairings emerging from a phenomenological mechanism and symmetry analysis}

To analyze the possible anisotropic spin singlet pairings for Bi$_2$X$_3$, we first identify all the possible anisotropic spin singlet pairing channels in a general manner from symmetry considerations and then analyze their properties. The actual relevance and the most probable channel of the anisotropic spin singlet pairings are to be presented in the next section. So, we consider in this section a realistic model for the band structure and a phenomenological correlation term that supports only spin singlet pairings. The model thus consists of two terms $H=H_{0}+H_{ex}$, in which $H_{0}$ is a tight binding term giving rise to the normal state band structure and $H_{ex}$ is an exchange term that could give rise to the desired spin singlet pairings.

Bi$_2$X$_3$ materials belong to the $D_{3d}^{5}$ space group,
which consists of Bi$_2$X$_3$ quintuple layers stacked along the out of plane direction.
Former theoretical studies on this material system are mostly based on two orbital $\mathbf{k}\cdot\mathbf{p}$ models
defined close to the $\boldsymbol{\Gamma}$ point.\cite{zhang09,liu10,fu09} To consider the
SC phase transition, we construct a tight binding model with the correct symmetry in the full BZ.\cite{dresselhaus08} Instead of working in the original BZ for a $D_{3d}^{5}$ space group\cite{zhang09,liu10}, we consider a hexagonal BZ corresponding to an
equivalent hexagonal lattice with two orbitals per unit cell, for its simplicity
and capability to respect the low energy symmetries and to account for physical properties of the system,
as verified in previous studies.\cite{wang10,hao11,sasaki11} Take the basis vector as $\phi_{\mathbf{k}}^{\dagger}=[a_{\mathbf{k}\uparrow}^{\dagger},
b_{\mathbf{k}\uparrow}^{\dagger},a_{\mathbf{k}\downarrow}^{\dagger},b_{\mathbf{k}\downarrow}^{\dagger}]$,
in which $a$ and $b$ operators correspond to the two orbitals, the tight binding model is (see Appendix A for details)
\begin{eqnarray}
H_{0}(\mathbf{k})&=&\epsilon(\mathbf{k})I_{4}+M(\mathbf{k})\Gamma_5+B_{0}c_{z}(\mathbf{k})\Gamma_{4}
+A_{0}[c_{y}(\mathbf{k})\Gamma_{1}   \notag \\
&&-c_{x}(\mathbf{k})\Gamma_{2}]+R_{1}d_{1}(\mathbf{k})\Gamma_{3}+R_{2}d_{2}(\mathbf{k})\Gamma_{4},
\end{eqnarray}
in which $\epsilon(\mathbf{k})=C_{0}+2C_{1}[1-\cos(\mathbf{k}\cdot\boldsymbol{\delta}_{4})]
+\frac{4}{3}C_{2}[3-\cos(\mathbf{k}\cdot\boldsymbol{\delta}_{1})-\cos(\mathbf{k}\cdot\boldsymbol{\delta}_{2})
-\cos(\mathbf{k}\cdot\boldsymbol{\delta}_{3})]$. $M(\mathbf{k})$ is obtained from $\epsilon(\mathbf{k})$
by making the substitutions $C_{i}\rightarrow M_{i} (i=0,1,2)$. $c_{x}(\mathbf{k})=\frac{1}{\sqrt{3}}
[\sin(\mathbf{k}\cdot\boldsymbol{\delta}_{1})-\sin(\mathbf{k}\cdot\boldsymbol{\delta}_{2})]$,
$c_{y}(\mathbf{k})=\frac{1}{3}[\sin(\mathbf{k}\cdot\boldsymbol{\delta}_{1})+\sin(\mathbf{k}\cdot\boldsymbol{\delta}_{2})
-2\sin(\mathbf{k}\cdot\boldsymbol{\delta}_{3})]$, and $c_{z}(\mathbf{k})=\sin(\mathbf{k}\cdot\boldsymbol{\delta}_{4})$.
Finally, $d_{1}(\mathbf{k})=-\frac{8}{3\sqrt{3}}[\sin(\mathbf{k}\cdot\mathbf{a}_{1})+\sin(\mathbf{k}\cdot\mathbf{a}_{2})
+\sin(\mathbf{k}\cdot\mathbf{a}_{3})]$ and $d_{2}(\mathbf{k})=-8[\sin(\mathbf{k}\cdot\boldsymbol{\delta}_{1})
+\sin(\mathbf{k}\cdot\boldsymbol{\delta}_{2})+\sin(\mathbf{k}\cdot\boldsymbol{\delta}_{3})]$. Here, the four independent nearest neighbor (NN) bond vectors of the effective hexagonal lattice are $\boldsymbol{\delta}_{1}=(\frac{\sqrt{3}}{2}a, \frac{1}{2}a, 0)$,
$\boldsymbol{\delta}_{2}=(-\frac{\sqrt{3}}{2}a$, $\frac{1}{2}a, 0)$, $\boldsymbol{\delta}_{3}=(0, -a, 0)$, and $\boldsymbol{\delta}_{4}=(0, 0, c)$, with $a$ and $c$ denoting in plane and out-of-plane lattice parameters.\cite{acparameters}
The three in plane second nearest neighbor (2NN) bond vectors in $d_{1}(\mathbf{k})$ are $\mathbf{a}_{1}=\boldsymbol{\delta}_{1}-\boldsymbol{\delta}_{2}$,
$\mathbf{a}_{2}=\boldsymbol{\delta}_{2}-\boldsymbol{\delta}_{3}$, and $\mathbf{a}_{3}=\boldsymbol{\delta}_{3}-\boldsymbol{\delta}_{1}$.
Expanding close to the $\boldsymbol{\Gamma}$ point, the above model is easily shown to reduce to the same form as Eqs. (16) and (17) in Liu \emph{et al}.\cite{liu10}
Demanding that the expanded model is the same as that in Liu \emph{et al}, the parameters are determined as shown in Table I.
In actual calculations, we change the value of $M_{1}$ to the bracketed value of 0.62 eV (0.102 eV) for Bi$_2$Se$_3$ (Bi$_2$Te$_3$) which yields a band gap of approximately 0.26 eV (0.06 eV).\cite{zhang09} All other parameters will be kept as given in Table I.

\begin{table}[ht]
\caption{Parameters in the tight binding model obtained by comparing with a $\mathbf{k}\cdot\mathbf{p}$ model\cite{liu10}, in units of electron volts. Bracketed values of $M_{1}$ are those actually used.} \centering
\begin{tabular}{c c c c c c}
\hline\hline
$$ & $C_{0}$ & $C_{1}$ & $C_{2}$ & $M_{0}$ & $M_{1}$\\ [0.2ex]
\hline
Bi$_2$Se$_3$ & -0.0083 & 0.063 & 1.774 & -0.28 & 0.0753 (0.62)\\
\hline
Bi$_2$Te$_3$ & -0.18 & 0.0634 & 2.59 & -0.3 & 0.027 (0.102)\\
\hline  
$$ & $M_{2}$ & $A_{0}$ & $B_{0}$ & $R_{1}$ & $R_{2}$ \\ [0.2ex]
\hline
Bi$_2$Se$_3$ & 2.596 & 0.804 & 0.237 & 0.713 & -1.597 \\
\hline
Bi$_2$Te$_3$ & 2.991 & 0.655 & 0.0295 & 0.536 & -1.064 \\
\hline
\hline
\end{tabular}
\end{table}

In Eq. (1), $I_4$ is the $4\times4$ unit matrix. The form of the $\Gamma$ matrices
depends on the choice of bases for the two orbitals.
In the original $\mathbf{k}\cdot\mathbf{p}$ model\cite{liu10,zhang09}, the two orbitals are chosen to have definite parity. Here, we choose a basis set in which the two orbitals have the physical meaning of local $p_z$ orbitals residing on the top and
bottom Se (or Te) layers of a quintuple unit hybridized with $p_z$ orbitals in neighboring Bi layers.\cite{liu10,wang10,hao11,fu09,fu10,sasaki11} The two basis sets are related by a simple unitary transformation.
We thus have $\Gamma_1=s_{1}\otimes\sigma_{3}$, $\Gamma_2=s_{2}\otimes\sigma_{3}$, $\Gamma_3=s_{3}\otimes\sigma_{3}$,
$\Gamma_{4}=-s_{0}\otimes\sigma_{2}$, and $\Gamma_{5}=s_{0}\otimes\sigma_{1}$.\cite{liu10,zhang09}
$s_{i}$ and $\sigma_{i}$ are Pauli matrices for the spin and orbital degrees of freedom.

To study possible spin singlet pairings, $H_{ex}$ is restricted to
contain only antiferromagnetic (AF) terms up to NN in plane bonds. The AF exchange
terms are known to be able to give rise to spin singlet pairings.\cite{seo08,kotliar88,goswami10}
Since there are two orbitals, we have both intraorbital and interorbital terms,
thus $H_{ex}=H_{intra}+H_{inter}$. They are written generally as
\begin{equation}
H_{intra}=\sum\limits_{\mathbf{i}\boldsymbol{\delta}\alpha}
J^{\alpha}_{\mathbf{i},\mathbf{i}+\boldsymbol{\delta}}(\mathbf{S}_{\mathbf{i}\alpha}\cdot\mathbf{S}_{\mathbf{i}+\boldsymbol{\delta},\alpha}
-\frac{1}{4}\hat{n}_{\mathbf{i}}^{\alpha}\hat{n}_{\mathbf{i}+\boldsymbol{\delta}}^{\alpha}),
\end{equation}
and
\begin{equation}
H_{inter}=\sum\limits_{\mathbf{i}\boldsymbol{\delta}}
J^{ab}_{\mathbf{i},\mathbf{i}+\boldsymbol{\delta}}
(\mathbf{S}_{\mathbf{i}a}\cdot\mathbf{S}_{\mathbf{i}+\boldsymbol{\delta},b}
-\frac{1}{4}\hat{n}_{\mathbf{i}}^{a}\hat{n}_{\mathbf{i}+\boldsymbol{\delta}}^{b}),
\end{equation}
where $\mathbf{i}$ runs over unit cells, $\boldsymbol{\delta}$ runs over the six NN in plane bonds $\pm\boldsymbol{\delta}_{j}$
($j$=1, 2, 3), and the $\alpha$ summation in $H_{intra}$ runs over the two orbitals.
$\hat{n}_{\mathbf{i}}^{\alpha}=\hat{n}_{\mathbf{i}\alpha\uparrow}+\hat{n}_{\mathbf{i}\alpha\downarrow}$
is the electron number operator for $\alpha$ orbital.

Out of all the possible spin singlet pairing channels contained in $H_{ex}$,
we focus on those pairings compatible with the crystal symmetry
of the Bi$_{2}$X$_{3}$ materials.
Taking advantage of the various irreducible
representations identified earlier by Liu \emph{et al}\cite{liu10},
we find six TRI $\mathbf{k}$-dependent pairings which
are compatible with the crystal symmetry and
are spin singlets (See Appendix C).
The six pairings identified include two intraorbital channels belonging
to the $\tilde{\Gamma}_{3}^{+}$ representation, which are
$\Delta_{j}\phi_{\mathbf{k}}^{\dagger}i\Gamma_{31}(\phi_{-\mathbf{k}}^{\dagger})^{\text{T}}\varphi_{j}(\mathbf{k})$
($j$=1, 2).
$\Delta_{j}$ are the magnitudes of the pairing terms.
$\Gamma_{31}=s_{2}\otimes\sigma_{0}$.
$\varphi_{1}(\mathbf{k})=\cos(\mathbf{k}\cdot\boldsymbol{\delta}_{1})-\cos(\mathbf{k}\cdot\boldsymbol{\delta}_{2}) =-2\sin(\frac{\sqrt{3}}{2}k_{x}a)\sin(\frac{1}{2}k_{y}a)$
and
$\varphi_{2}(\mathbf{k})=2\cos(\mathbf{k}\cdot\boldsymbol{\delta}_{3})-\cos(\mathbf{k}\cdot\boldsymbol{\delta}_{1})
-\cos(\mathbf{k}\cdot\boldsymbol{\delta}_{2})=2[\cos(k_{y}a)-\cos(\frac{\sqrt{3}}{2}k_{x}a)\cos(\frac{1}{2}k_{y}a)]$. There are also two
interorbital pairings belonging to the  $\tilde{\Gamma}_{3}^{+}$
representation, which could be obtained from the above two pairing
terms by substituting $\Gamma_{31}$ by
$\Gamma_{24}=s_{2}\otimes\sigma_{1}$ and identifying
$\varphi_{3(4)}(\mathbf{k})=\varphi_{1(2)}(\mathbf{k})$. The other
two pairing channels belong to $\tilde{\Gamma}_{3}^{-}$ and are
$\Delta_{j}\phi_{\mathbf{k}}^{\dagger}i\Gamma_{25}(\phi_{-\mathbf{k}}^{\dagger})^{\text{T}}\varphi_{j}(\mathbf{k})$
($j$=5, 6), in which $\Gamma_{25}=s_{2}\otimes\sigma_{2}$,
$\varphi_{5}(\mathbf{k})=\sin(\mathbf{k}\cdot\boldsymbol{\delta}_{1})-\sin(\mathbf{k}\cdot\boldsymbol{\delta}_{2}) =2\sin(\frac{\sqrt{3}}{2}k_{x}a)\cos(\frac{1}{2}k_{y}a)$
and
$\varphi_{6}(\mathbf{k})=\sin(\mathbf{k}\cdot\boldsymbol{\delta}_{1})+\sin(\mathbf{k}\cdot\boldsymbol{\delta}_{2})
-2\sin(\mathbf{k}\cdot\boldsymbol{\delta}_{3})=2\sin(\frac{1}{2}k_{y}a)[\cos(\frac{\sqrt{3}}{2}k_{x}a)+2\cos(\frac{1}{2}k_{y}a)]$.
In the following discussions, we use $\Delta_j$ to refer to the $j$-th pairing defined above, in places where confusion is not incurred.

The wave vector dependencies of the various pairing components determine their gap structures. For experimentally relevant chemical potentials, states on the Fermi surface are all close to the $\boldsymbol{\Gamma}=(0,0,0)$ point of the BZ, it is thus enough to focus on small wave vectors. For the first and the third pairings, $\varphi_{1}(\mathbf{k})=0$ gives two line nodes of the gap along $k_{x}=0$ and $k_{y}=0$. The other nodes determined by $k_{x}=\frac{2\pi}{\sqrt{3}a}$ or $k_{y}=\frac{2\pi}{a}$ are unlikely to occur since they lie on the BZ boundary and are far away from the Fermi surface. For the second and the fourth pairings, the line nodes are determined by $\varphi_{2}(\mathbf{k})=0$ and satisfy $4\cos(\frac{1}{2}k_{y})=\cos(\frac{\sqrt{3}}{2}k_{x})+\alpha\sqrt{\cos^{2}(\frac{\sqrt{3}}{2}k_{x})+8}$, $\alpha=\pm$. Note that, for each $k_{x}$, there are two solutions for $k_{y}$ for each $\alpha$. Usually, only the two solutions related to $\alpha=+$ lie on the Fermi surface, so the second and fourth pairings in general also have two lines nodes. Similarly, from $\varphi_{5}(\mathbf{k})=0$ we know that the fifth pairing usually has only one line node determined by states on the Fermi surface with $k_{x}=0$. And finally the sixth pairing also has only one set of line nodes determined by $\varphi_{6}(\mathbf{k})=0$, consisting of states on the Fermi surface with $k_{y}=0$.
Among the six pairings identified, the fifth ($\Delta_{5}$) and the sixth ($\Delta_{6}$)
pairings are peculiar in that, though spin singlet, they have \emph{$p$-wave like
odd $\mathbf{k}$ dependencies} for small wave vectors, which together with their odd orbital-parity
is consistent with their spin singlet nature.   
Here, we define spatial-parity and orbital-parity as the parities of the SC
order parameter related with reversal of the wave vector and exchange of the two orbitals,
respectively.

To check properties of the pairings identified above,
we first calculate their surface local density of states (SLDOS), which
are directly observable via point contact
spectra.\cite{sasaki11,koren11,kirzhner12} The SLDOS are defined as
the surface spectral function averaged over the surface wave vectors,
that is
$\rho_{s}(\omega)=N_{s}^{-1}\sum_{\mathbf{k}_{xy}}A(\mathbf{k}_{xy},\omega)$,
in which $N_{s}$ is the number of wave vectors $\mathbf{k}_{xy}$ in the
surface BZ.\cite{sasaki11} $A(\mathbf{k}_{xy},\omega)$ is the
surface spectral function defined as imaginary part of the
electronic surface Green's functions (GF) obtained in terms of the
iterative GF method (or, transfer matrix method).\cite{hao11,wang10} See Appendix D for a brief explanation of our usage of the iterative GF method. To distinguish possible
surface-state contributions, we calculate simultaneously the bulk local
density of states (BLDOS), which are obtained easily from the bulk GF.
In these calculations, we
first add directly a certain pairing term to $H_{0}$ without regarding its origin, to focus on a
single pairing channel.

\begin{figure}
\centering
\includegraphics[width=8.8cm,height=12cm,angle=0]{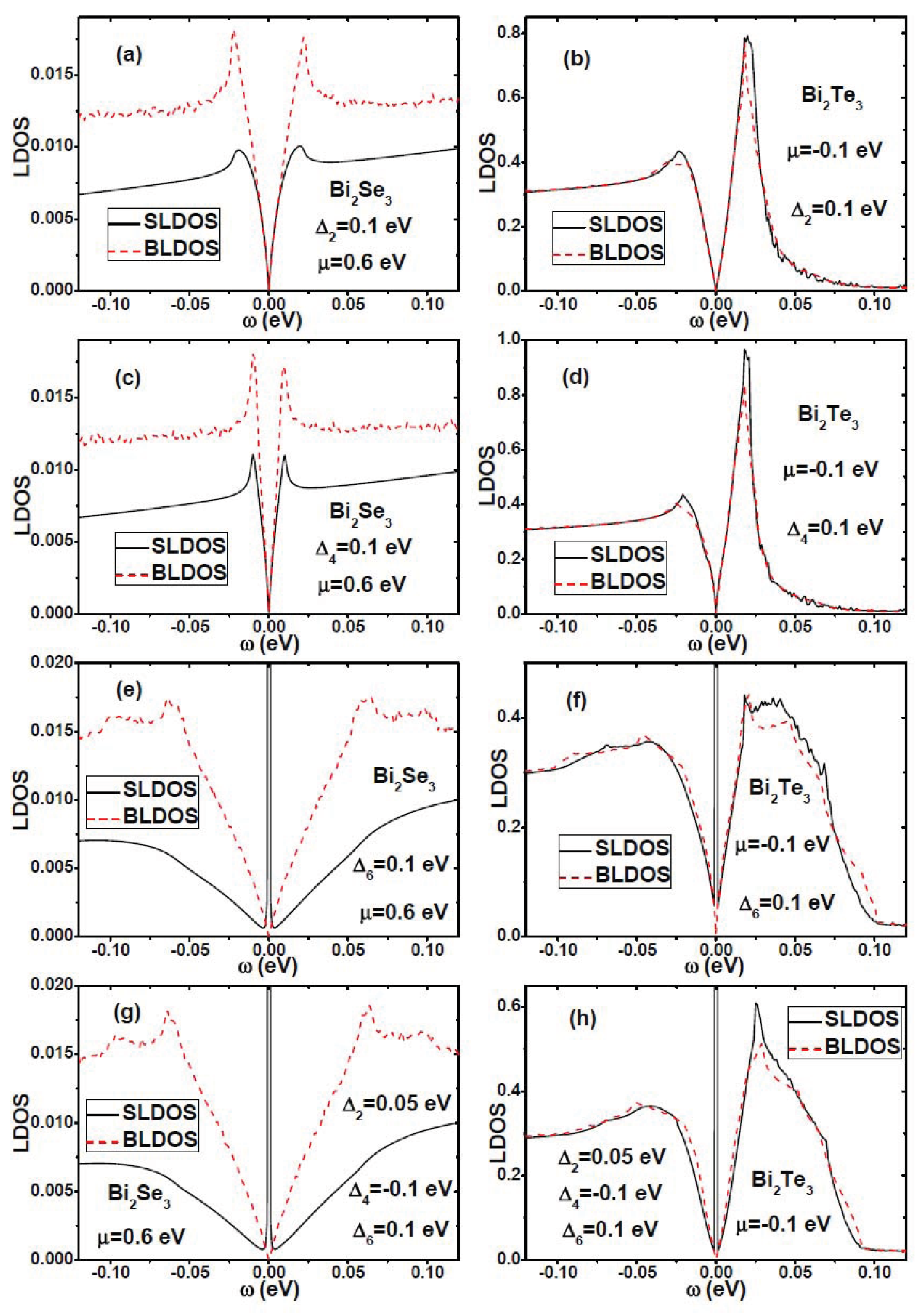}
\caption{The surface (SLDOS) and bulk (BLDOS) local density of states,
for three typical SC pairings that could possibly realize
in Bi$_2$Se$_3$ (a, c, e) and Bi$_2$Te$_3$ (b, d, f). (g) and (h) show results for representative
mixed pairing states. The chemical potential and nonzero pairing components and their amplitudes are as indicated.}
\end{figure}

As shown in Figs. 1(a) to 1(f) are the LDOS for three
typical pairings, for Bi$_{2}$Se$_{3}$ (Bi$_{2}$Te$_{3}$) system at
electron (hole) fillings specified by the chemical potentials as
indicated. For the same pairing, qualitatively similar results are
obtained for the two systems.
BLDOS for the $\Delta_{1}$ to $\Delta_{4}$
pairings all show a V-shape demonstrating the linear density of states at the chemical potential.
The SLDOS for these pairings are all similar to the BLDOS, indicating that
there are no topological nontrivial surface ABS.
Very interestingly, in Figs. 1(e) and 1(f) which are typical results
for both $\Delta_{5}$ and
$\Delta_{6}$ pairings, while the BLDOS still shows a linear density of states,
a sharp zero energy peak appears in the SLDOS.
This peak structure is reminiscent of the ZBCP observed
in some point contact spectra measurements in superconducting
Cu$_{x}$Bi$_{2}$Se$_{3}$.\cite{sasaki11,kirzhner12} While these zero energy peaks imply sharp ZBCPs in point contact spectra and STM experiments, they would broaden at finite temperature and thus within the experimental resolution.\cite{sasaki11,hsieh12}
We also find that, a TRI pairing consisting of a mixture of $\Delta_{5}$ or
$\Delta_{6}$ and other pairings preserves the novel zero energy peak in the SLDOS,
see Figs. 1(g) and 1(h) for typical examples. Note that, because all the six pairings share the gap node at $k_{x}=k_{y}=0$, the composite pairing consisting of several pairing components is still gapless in the bulk. This explains the V-shape BLDOS shown in Figs.1(g) and 1(h).

The surface ABS are more clearly seen from the surface spectral functions,
as shown in Fig. 2 for $\Delta_{6}$.
Along ($k_{x}$, 0) direction of the surface BZ, the superconductor has a line node. Along other
directions such as (0, $k_{y}$) in Fig. 2, a SC gap opens,
the surface ABS are clearly present and form a flat band.
This is similar to the surface ABS in some nodal spin triplet pairings.\cite{sasaki11,hao11}
Though the topological numbers for $\Delta_5$ and $\Delta_6$ with nodal lines are difficult to
calculate directly \cite{sato10,sasaki11,matsuura13}, the number counting of zero energy surface ABS are also indicative
of important conclusions. By explicitly calculating the eigenstates for a thin film
of the $\Delta_{6}$ pairing, we confirmed that a \emph{single} Kramers' pair of ABS exists
on both the top and the bottom surfaces.
Since an odd number of Kramers' pairs of surface states are generally protected from TRI perturbations,
the gapless surface ABS for $\Delta_5$ and $\Delta_6$ should also be topologically stable.

\begin{figure}
\centering
\includegraphics[width=8.5cm,height=12.5cm,angle=0]{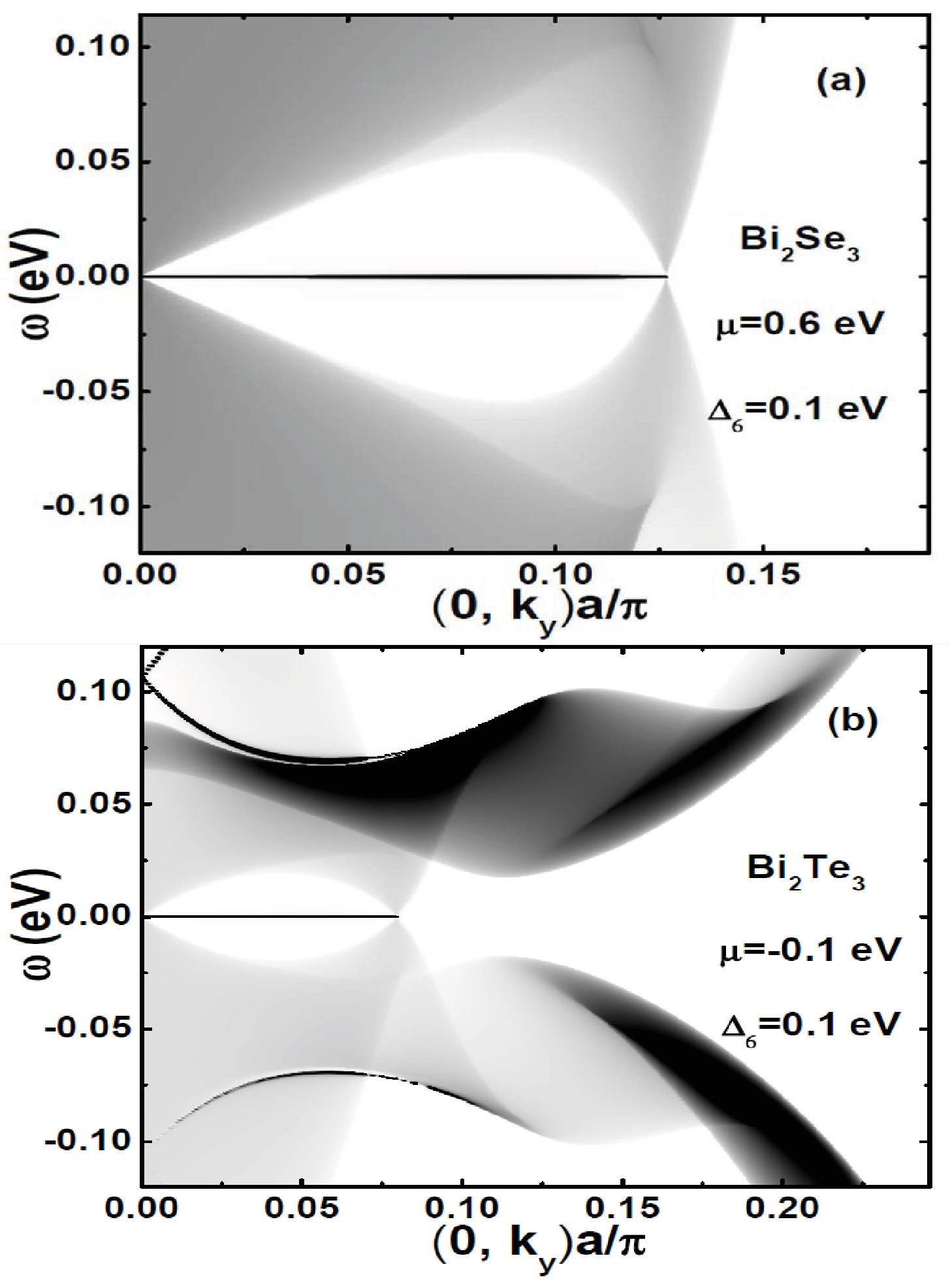}
\caption{The surface spectral function for the sixth pairing identified in this work,
for (a) electron doped Bi$_{2}$Se$_{3}$ and (b) Bi$_{2}$Te$_{3}$ with Fermi surface
crossing the valence band. The darker the color, the larger the spectral weight.}
\end{figure}

\section{\label{sec:result2}pairings emerging from purely repulsive Coulomb interactions}

Before studying further properties of the novel spin singlet pairings identified above,
it is important to ask if the spin singlet pairings are actually relevant
to the superconducting state of Bi$_2$X$_3$ (X is Se or Te).
That is, whether or not spin singlet pairing
is the dominant pairing channel for realistic pairing mechanisms.
Since evidences have appeared that Bi$_2$Se$_3$ is a sizable correlated electron
system \cite{wang11,craco12}, whereas the electron-phonon
coupling (EPC) in the present system is generally considered to be smaller than BCS
superconductors\cite{giraud11,hatch11,zhu12,pan12}, we would here take the \emph{purely repulsive}
short range Coulomb interaction as the pairing mechanism for Bi$_2$X$_3$.
Since EPC usually favors spin singlet pairings, the following study stands
as a more stringent test for the relevance of spin singlet pairings.

Compatible with our orbital convention, the Coulomb repulsion terms
are conveniently added to $H_{0}$ as
\begin{equation}
H_{1}=U\sum\limits_{\mathbf{i}\alpha}\hat{n}_{\mathbf{i}\alpha\uparrow}\hat{n}_{\mathbf{i}\alpha\downarrow}
+V\sum\limits_{\mathbf{i}}\hat{n}_{\mathbf{i}}^{a}\hat{n}_{\mathbf{i}}^{b},
\end{equation}
in which $\alpha$ runs over the two orbitals.
Since the two orbitals in the present model reside on different sites within a unit cell,
we do not include the Hund's coupling term between them.\cite{craco12} In addition, we expect
the on-site intraorbital correlation stronger than the inter-site interorbital
correlation, that is $U>V>0$.

Starting from $H_{0}+H_{1}$, the dominant superexchange couplings
are derived using standard projection operator method
(Schrieffer-Wolf transformation).\cite{schrieffer66}
Here, we retain the lowest order terms up to two site correlations.
The superexchange terms are derived first at half filling. The relation $U>V$
imposes a local intraorbital no-double-occupancy condition.
In principle, each hopping parameter and every pair of two different hopping parameters listed in Table I could mediate
a superexchange term. However, since the magnitudes of the various parameters differ greatly,
we expect that the superexchange terms mediated by the largest several parameters
in Table I dominate the actual pairing instability. We thus select $C_2$, $M_2$ and $R_2$, which are apparently
larger than the other parameters.
We find that $C_2$ mediates the term $H_{intra}$,
with $J^{\alpha}_{\mathbf{i},\mathbf{i}+\boldsymbol{\delta}}=\frac{8C^{2}_{2}}{9U}$ independent of $\boldsymbol{\delta}$.
$M_2$, $R_2$ and their crossing mediate three interorbital superexchange
terms, which combine to give $H_{inter}$, with $J^{ab}_{\mathbf{i},\mathbf{i}+\boldsymbol{\delta}}=
\frac{(8R_{2})^2}{U}[1+(-1)^{\eta(\boldsymbol{\delta})}\frac{M_{2}}{6R_{2}}]^{2}$.
$\eta(\boldsymbol{\delta})$ arises from the crossing term
between $M_{2}$ and $R_{2}$ and is defined as $0$ ($1$) for
$\boldsymbol{\delta}=\boldsymbol{\delta}_{j}$
($\boldsymbol{\delta}=-\boldsymbol{\delta}_{j}$), with $j$=1, 2, 3.
Thus, the AF exchange terms conceived in Eqs. (2)-(3) emerge naturally
in Bi$_2$X$_3$ if we consider short range
repulsive Coulomb interaction as the pairing mechanism.
Other parameters, such as
$A_{0}$ in Table I which could mediate
superexchange correlations favoring triplet pairings, are too small to be competitive with the
identified terms, for both Bi$_{2}$Se$_{3}$ and Bi$_{2}$Te$_{3}$. More details about the derivation of the dominant superexchange terms can be found in Appendix B. In conclusion, pairings mediated by purely repulsive short range Coulomb interactions
are dominantly spin singlet in Bi$_2$X$_3$.

Besides confirming the hypothetical form of $H_{ex}$, an interesting new consequence
of the above derivation is that $H_{inter}$ only has $C_3$
symmetry with respect to $c$-axis, which is clear from the $\boldsymbol{\delta}$-dependency of
$\eta(\boldsymbol{\delta})$.
This feature inherits directly from the $R_{2}$ term, because the
crossing term between $R_{2}$ and $M_{2}$ is linear in $R_{2}$.
An immediate consequence of this real space
anisotropic correlation is that it explicitly \emph{breaks the in
plane inversion symmetry} of $H_{ex}$, which implies that
the \emph{in plane spatial-parity} is not a good quantum number and the resulting pairing
would be a mixture of even and odd spatial-parity states.
The spatial-parity mixing
effect is more clearly seen from the Fourier transformation of the
interorbital pairing potential in $H_{inter}$, which is
\begin{eqnarray}
J^{ab}_{\mathbf{k},\mathbf{k'}}&=&\frac{(8R_{2})^2}{NU}\sum\limits_{j=1}^{3}\{[1+\frac{M_{2}^{2}}{(6R_{2})^2}]
\cos(\mathbf{k}-\mathbf{k'})\cdot\boldsymbol{\delta}_{j}  \notag \\
&&+i\frac{M_{2}}{3R_{2}}
\sin(\mathbf{k}-\mathbf{k'})\cdot\boldsymbol{\delta}_{j}\},
\end{eqnarray}
with $N$ the number of unit cells in the whole lattice.
This dynamical generation of an inversion symmetry breaking \emph{correlation} term is
an essential feature of the present model and is intrinsic to
Bi$_{2}$X$_{3}$ materials. We also mention that, though the $R_2$ term in $H_{0}$ already
breaks the in-plane inversion, the present correlation term is different because firstly
it depends also on $M_2$ and emerges as a crossing term between the $M_2$ and $R_2$ terms,
secondly the mixing of pairings with even and odd spatial-parity is now explicit
and their relative phase and weight are determined by the correlation term itself.

To see the relative importance of the six pairings identified formerly
by symmetry, we have performed mean field calculations at zero
temperature, in terms of a $t$$-$$U$$-$$V$$-$$J$ type full model $H=H_{0}+H_{1}+H_{ex}$ (see Appendix C for details).
First, we decouple $H_{ex}$
by introducing nonequivalent bond pairing terms.\cite{seo08,goswami10,kotliar88}
Six intraorbital pairings for the $a$ orbital are introduced as $\chi_{a}^{\nu\pm}=\langle
a_{\mathbf{i}\pm\boldsymbol{\delta}_{\nu},\downarrow}a_{\mathbf{i}\uparrow}
-a_{\mathbf{i}\pm\boldsymbol{\delta}_{\nu},\uparrow}a_{\mathbf{i}\downarrow}\rangle$
($\nu=1, 2, 3$). Another six intraorbital pairings
$\chi_{b}^{\nu\pm}$ are similarly defined for the $b$ orbital.
Six interorbital pairings are introduced as
$\chi_{ba}^{\nu\pm}=\langle
b_{\mathbf{i}\pm\boldsymbol{\delta}_{\nu},\downarrow}a_{\mathbf{i}\uparrow}
-b_{\mathbf{i}\pm\boldsymbol{\delta}_{\nu},\uparrow}a_{\mathbf{i}\downarrow}\rangle$
($\nu=1, 2, 3$). A translational invariant pairing phase is
assumed, so that the $18$ pairing terms are independent of
$\mathbf{i}$. Since we focus on the SC phase, the
decoupling of $H_{ex}$ to the normal phase is disregarded.
Terms in $H_{1}$ are decoupled in the simplest manner as
$\hat{n}_{1}\hat{n}_{2}\rightarrow\langle\hat{n}_{1}\rangle\hat{n}_{2}+\hat{n}_{1}\langle\hat{n}_{2}\rangle
-\langle\hat{n}_{1}\rangle\langle\hat{n}_{2}\rangle$.
For each set of parameters ($U$, $V$ and doping or chemical potential), we then get the $18$ mean field pairing order
parameters self-consistently from many different initial values.
Finally, the amplitudes of the six pairing terms are projected out of
the solution.\cite{seo08,goswami10,kotliar88}
We find that, for parameters typical for
Bi$_{2}$Se$_{3}$ and Bi$_{2}$Te$_{3}$, the dominant pairing
channel is a mixture of $\Delta_{4}$ and $\Delta_{6}$ pairings with
a tiny admixture of the $\Delta_{2}$ pairing. All other pairing components are
identically zero. In addition, though there are usually several
coexisting pairing components, the pairing is \emph{time reversal invariant} up to
a global $U(1)$ phase, same as the conclusion of a similar calculation
for iron pnictides.\cite{seo08} We emphasize that the two main features of the results, coexistence of several pairing components and the time reversal invariance of the full pairing, are true for all model parameters that we have tried, which are spanned by $U\sim[8$ eV, 20 eV], $V\sim[5$ eV, 8 eV] and $x\sim[-0.08, 0.16]$ ($x$ is the number of excess electrons in each unit cell). The robustness of the two features are consequences of the interorbital superexchange correlation term, as shown in Eq.(5), from which the pairings of even spatial-parity and of odd spatial-parity always appear together in a time reversal invariant combination. We should confess that the present mean field studies underestimate the
fluctuation effect and the competition from possible magnetic normal states,
so that the pairing instability is overestimated. However, the dominant
pairing channel should still be spin singlet even if these corrections
are taken into account, which are left to future works.

\section{\label{sec:Summary}Experimental implications and Summary}

The novel spin singlet pairings, $\Delta_{5}$ and $\Delta_{6}$, could be distinguished experimentally from other candidate pairings. Firstly, since they give quite different surface spectral functions\cite{hao11,sasaki11}, the correct pairing symmetry could be read from ARPES if the precision of measurement can reach the order of $\sim0.1$ meV.\cite{wray10} Secondly, the SLDOS which could be probed by point contact spectroscopy or STM can also be used to discriminate among the candidate pairings. For example, the fully gapped interorbital triplet pairing gives an in-gap nonzero energy double-peak structure in SLDOS at low temperature and for good contact.\cite{sasaki11,hsieh12} However, the $\Delta_{5}$ or $\Delta_{6}$ pairing always gives a single ZBCP. While our proposal is in better agreement with existing experiments\cite{sasaki11,koren11,kirzhner12,chen12}, more measurements are desired to get a definite conclusion. Thirdly, the static spin susceptibilities show clear differences for different candidate pairings and thus could be used to discriminate some of them. Fourthly, the thermal conductivity which depends sensitively on the anisotropy of the pairings was proposed to discriminate two triplet pairings.\cite{nagai12} It should also be able to tell $\Delta_5$ or $\Delta_6$ pairing from the other candidate (triplet) pairings. Details that lead to the above conclusions are to be published elsewhere. Besides the above proposals, our spin singlet pairing state is not in direct contradiction with existing experiments. Not only for experiments pointing to polar or anisotropic pairings, it could also be in agreement with a specific heat experiment which shows that the pairing has a fully gapped component.\cite{kriener11} Though the novel singlet pairings $\Delta_5$ or $\Delta_6$ are both gapless, a fully gapped $\mathbf{k}$-independent interorbital spin singlet component could be readily added into our mixed pairing state to give the experimental feature (see Appendix B).

To summarize, we have studied the possible anisotropic spin singlet pairings in Bi$_2$X$_3$ (X is Se or Te). Two novel interorbital spin singlet pairings with odd spatial-parity and odd orbital-parity support surface ABS, which form zero energy flat bands. The presence of only one Kramers' pair of ABS on each surface implies that they should be topologically stable against TRI perturbations. Considering purely repulsive short range Coulomb interaction as the pairing mechanism, the low energy effective model turns out indeed to be dominated by spin-singlet-favoring AF correlations. Besides, the interorbital AF correlation favors a pairing state with mixed spatial-parity. It would be interesting to see if this prediction can be verified by future experiments.

\begin{acknowledgments}
L.H. and G.L.W. are supported by NSFC.11204035 and SRFDP.20120092120040. T.K.L. acknowledges the support of NSC in Taiwan under Grant No.103-2120-M-001-009. J.W. is supported by NSFC.11274059 and NSF of Jiangsu Province BK20131284. W.F.T. is supported by the NSC in Taiwan under Grant No.102-2112-M-110-009. Part of the calculations was performed in the National Center for High-Performance Computing in Taiwan.
\end{acknowledgments}\index{}

\begin{appendix}

\section{\label{sec:TBModel}Tight Binding Model}

Here, we construct a tight binding model for the bulk electronic structures of Bi$_2$Se$_3$, Cu$_x$Bi$_{2}$Se$_3$ and Bi$_2$Te$_3$ materials from symmetry considerations. As illustrated in the main text, we replace the actual lattice with $D_{3d}^{5}$ space group symmetry by a hexagonal lattice with two orbitals per unit cell. Take in-plane (labeled as the $xy$ plane) and out-of-plane (labeled as the $z$ direction) lattice parameters as $a$ and $c$, the four independent nearest-neighbor (NN) bond vectors of the effective hexagonal lattice are $\boldsymbol{\delta}_{1}=(\frac{\sqrt{3}}{2}a, \frac{1}{2}a, 0)$,
$\boldsymbol{\delta}_{2}=(-\frac{\sqrt{3}}{2}a$, $\frac{1}{2}a, 0)$, $\boldsymbol{\delta}_{3}=(0, -a, 0)$, and $\boldsymbol{\delta}_{4}=(0, 0, c)$. We take the lattice parameters as $a$=4.14 \AA (4.38 \AA) and $3c$=28.64 \AA (30.487 \AA)
for Bi$_2$Se$_3$ (Bi$_2$Te$_3$).\cite{acparameters} Small changes in $a$ and $c$ for the SC state of Bi$_2$X$_3$ are neglected.\cite{hor10,wray10,zhang11pa,zhang11pb}
For Bi$_2$Te$_3$, we consider the SC transition under ambient pressure without structural transition and so the symmetry keeps as $D_{3d}^{5}$.\cite{zhang11pb}

Since spin-orbit interaction is important in Bi$_2$X$_3$, we have to consider the double group of the $D_{3d}^{5}$ space group to get a proper tight binding model. Following the notations of Liu \emph{et al} \cite{liu10}, we write the generators of the point group for $D_{3d}^{5}$ as $R_{3}$ (threefold rotation, about the symmetry line parallel to $z$ axis), $R_{2}$ (twofold rotation, about the symmetry line parallel to $x$ axis) and $P$ (inversion). For the double group, introduce the operator $\mathcal{C}$ to represent $2\pi$ rotation. The characters for the various irreducible representations are then as shown in Table I.\cite{dresselhaus08,liu10}

\begin{table*}[ht] 
\caption{Character table for the double group of $D_{3d}^{5}$ ($R\bar{3}m$).\cite{dresselhaus08,liu10}} \centering
\begin{tabular}{c c c c c c c c c c c c c}
\hline\hline \\ [-1.5ex]
$D_{3d}(\bar{3}m)$ & $E$ & 2$R_{3}$ & 3$R_{2}$ & $P$ & 2$P$$R_{3}$ & 3$P$$R_{2}$ & $\mathcal{C}$ & 2$\mathcal{C}$$R_{3}$ & 3$\mathcal{C}$$R_{2}$ & $\mathcal{C}$$P$ & 2$\mathcal{C}$$P$$R_{3}$ & 3$\mathcal{C}$$P$$R_{2}$   \\ [0.2ex]
\hline  \\ [-2ex]
$\tilde{\Gamma}_{1}^{+}$ & 1 & 1 & 1 & 1 & 1 & 1 & 1 & 1 & 1 & 1 & 1 & 1 \\
\hline  \\ [-2ex]
$\tilde{\Gamma}_{2}^{+}$ & 1 & 1 & -1 & 1 & 1 & -1 & 1 & 1 & -1 & 1 & 1 & -1 \\
\hline  \\ [-2ex]
$\tilde{\Gamma}_{3}^{+}$ & 2 & -1 & 0 & 2 & -1 & 0 & 2 & -1 & 0 & 2 & -1 & 0 \\
\hline  \\ [-2ex]
$\tilde{\Gamma}_{4}^{+}$ & 1 & -1 & $i$ & 1 & -1 & $i$ & -1 & 1 & $-i$ & -1 & 1 & $-i$ \\
\hline  \\ [-2ex]
$\tilde{\Gamma}_{5}^{+}$ & 1 & -1 & $-i$ & 1 & -1 & $-i$ & -1 & 1 & $i$ & -1 & 1 & $i$ \\
\hline  \\ [-2ex]
$\tilde{\Gamma}_{6}^{+}$ & 2 & 1 & 0 & 2 & 1 & 0 & -2 & -1 & 0 & -2 & -1 & 0 \\
\hline  \\ [-2ex]
$\tilde{\Gamma}_{1}^{-}$ & 1 & 1 & 1 & -1 & -1 & -1 & 1 & 1 & 1 & -1 & -1 & -1 \\
\hline  \\ [-2ex]
$\tilde{\Gamma}_{2}^{-}$ & 1 & 1 & -1 & -1 & -1 & 1 & 1 & 1 & -1 & -1 & -1 & 1 \\
\hline  \\ [-2ex]
$\tilde{\Gamma}_{3}^{-}$ & 2 & -1 & 0 & -2 & 1 & 0 & 2 & -1 & 0 & -2 & 1 & 0 \\
\hline  \\ [-2ex]
$\tilde{\Gamma}_{4}^{-}$ & 1 & -1 & $i$ & -1 & 1 & $-i$ & -1 & 1 & $-i$ & 1 & -1 & $i$ \\
\hline  \\ [-2ex]
$\tilde{\Gamma}_{5}^{-}$ & 1 & -1 & $-i$ & -1 & 1 & $i$ & -1 & 1 & $i$ & 1 & -1 & $-i$ \\
\hline  \\ [-2ex]
$\tilde{\Gamma}_{6}^{-}$ & 2 & 1 & 0 & -2 & -1 & 0 & -2 & -1 & 0 & 2 & 1 & 0 \\
\hline
\hline
\end{tabular}
\end{table*}

Same as in the main text, we take the basis vector as $\phi_{\mathbf{k}}^{\dagger}=[a_{\mathbf{k}\uparrow}^{\dagger},
b_{\mathbf{k}\uparrow}^{\dagger},a_{\mathbf{k}\downarrow}^{\dagger},b_{\mathbf{k}\downarrow}^{\dagger}]$,
in which the two orbitals represented by the $a$ (not to be confused with the in-plane lattice parameter) and $b$ operators denote local $p_z$ orbitals residing on the top and bottom Se (Te) layers of a Bi$_2$Se$_3$ (Bi$_2$Te$_3$) quintuple unit hybridized with $p_z$ orbitals in neighboring Bi layers. The fact that a minimal model consisting of the above two hybridized $p_z$ orbitals is enough for the low energy physics of topological insulators like Bi$_2$Se$_3$ and Bi$_2$Te$_3$ has been established in previous works.\cite{liu10,zhang09,fu09} With this basis at hand, and define $s_{i}$ and $\sigma_{i}$ as Pauli matrices for the spin and orbital degrees of freedom, we can write the various symmetry operations in matrix form.\cite{liu10} The time reversal operator is $T=is_{2}\otimes\sigma_{0}K$, where $K$ denotes the complex conjugation and $\sigma_{0}$ is the $2\times2$ unit matrix in orbital subspace. Matrix for the threefold rotation is $R_{3}=e^{i(s_{3}\otimes\sigma_{0}/2)\theta}=\cos{\frac{\theta}{2}}+is_{3}\otimes\sigma_{0}\sin{\frac{\theta}{2}}$, with $\theta=2\pi/3$. The twofold rotation is $R_{2}=is_{1}\otimes\sigma_{1}$. The matrix for inversion is $P=s_{0}\otimes\sigma_{1}$, with $s_{0}$ the $2\times2$ unit matrix in spin subspace. Finally, we may also write out the matrix for $\mathcal{C}$, the $2\pi$ rotation, which should be written as $-s_{0}\otimes\sigma_{0}$.

Denote by $H_{0}(\mathbf{k})$ the 4$\times$4 Hamiltonian matrix for wave vector $\mathbf{k}$ in the basis of $\phi_{\mathbf{k}}^{\dagger}$. The translational invariance (or, periodicity) of the material in real space implies that $H_{0}(\mathbf{k})$ is a periodic function of the reciprocal lattice vectors in the extended zone scheme.\cite{dresselhaus08} According to Bloch's theorem, $H_{0}(\mathbf{k})$ could be Fourier expanded in terms of the real space lattice. Since the reciprocal lattice have the same symmetry as the real space lattice, $H_{0}(\mathbf{k})$ should be an invariant under the action of the $D_{3d}^{5}$ double group. So, the general form of $H_{0}(\mathbf{k})$ conforming to symmetry is \cite{dresselhaus08}
\begin{equation}
H_{0}(\mathbf{k})=\sum\limits_{j,\alpha\nu}a_{j\alpha\nu}g_{j\alpha}(\mathbf{k},\mathbf{d}_{\nu})O^{j\alpha},
\end{equation}
where $\mathbf{d}_{\nu}$ represent lattice vectors connecting sites at the $\nu$-th nearest-neighbor, $g_{j\alpha}(\mathbf{k},\mathbf{d}_{\nu})$ is symmetrized combination of Fourier functions of the form $e^{i\mathbf{k}\cdot\mathbf{d}_{\nu}}$ which transforms as the $\alpha$-th component of the $j$-th irreducible representation of $D_{3d}^{5}$ double group, $O^{j\alpha}$ is a basis matrix function also transforming as the $\alpha$-th component of the $j$-th irreducible representation of the symmetry group, and $a_{j\alpha\nu}$ is a constant indicating the contribution of this term to the band structure. The above form is uniquely determined by group theory. Once the energy bands at high symmetry points of the BZ is known from experiments or first principle calculations, the coefficients $\{a_{j\alpha\nu}\}$ could be fixed by a minimization procedure. The model determined by the above Slater-Koster method is naturally a tight binding model if the number of relevant $\mathbf{d}_{\nu}$ is finite and $a_{j\alpha\nu}$ for further neighbor contributions are negligible.

For the materials of interest to us here, the most reliable and detailed data about the low energy band structure is around the BZ center, that is the $\boldsymbol{\Gamma}$ point.\cite{liu10} Since the properties close to the $\boldsymbol{\Gamma}$ point is of most interest to us, this information though insufficient to obtain a model which could produce the correct complete band structure should still be able to provide us a model of the correct global symmetry with correct behavior close to the $\boldsymbol{\Gamma}$ point. On the other hand, we suppose that a tight binding model (TBM) is sufficient to give a good description of the electronic band structure, in terms of $\mathbf{d}_{\nu}$ as short ranged as possible.

The basis matrix functions are formerly constructed by Liu \emph{et al}.\cite{liu10} They are cited here as the second column of Table II. Their time reversal properties are cited as the third column. In our orbital convention, the $\Gamma$ matrices are defined as $\Gamma_1=s_{1}\otimes\sigma_{3}$, $\Gamma_2=s_{2}\otimes\sigma_{3}$, $\Gamma_3=s_{3}\otimes\sigma_{3}$, $\Gamma_{4}=-s_{0}\otimes\sigma_{2}$, $\Gamma_{5}=s_{0}\otimes\sigma_{1}$, and $\Gamma_{ij}=[\Gamma_{i}, \Gamma_{i}]/2i$ are commutators of corresponding $\Gamma_{i}$.\cite{liu10,zhang09} In addition, $I_{4}$ is the 4$\times$4 unit matrix.

\begin{table*}[ht] 
\caption{Basis matrix functions (second column) in terms of the $\Gamma$ matrices\cite{liu10}, and the symmetrized Fourier functions (fourth column).} \centering
\begin{tabular}{c c c c c}
\hline\hline \\ [-1.5ex]
Representation & Basis matrices & $T$ & Basis Fourier Functions & $T'$ \\ [0.2ex]
\hline  \\ [-2ex]
$\tilde{\Gamma}_{1}^{+}$ & $I_{4}$ & $+$ & $1, \frac{1}{3}[\cos(\mathbf{k}\cdot\boldsymbol{\delta}_{1})+\cos(\mathbf{k}\cdot\boldsymbol{\delta}_{2})+\cos(\mathbf{k}\cdot\boldsymbol{\delta}_{3})], \cos(\mathbf{k}\cdot\boldsymbol{\delta}_{4})$ & $+$ \\
\hline  \\ [-2ex]
$\tilde{\Gamma}_{1}^{+}$ & $\Gamma_{5}$ & $+$ & $1, \frac{1}{3}[\cos(\mathbf{k}\cdot\boldsymbol{\delta}_{1})+\cos(\mathbf{k}\cdot\boldsymbol{\delta}_{2})+\cos(\mathbf{k}\cdot\boldsymbol{\delta}_{3})], \cos(\mathbf{k}\cdot\boldsymbol{\delta}_{4})$ & $+$  \\
\hline  \\ [-2ex]
$\tilde{\Gamma}_{2}^{+}$ & $\Gamma_{12}$ & $-$ & none & none \\
\hline  \\ [-2ex]
$\tilde{\Gamma}_{2}^{+}$ & $\Gamma_{34}$ & $-$ & none & none \\
\hline  \\ [-2ex]
$\tilde{\Gamma}_{3}^{+}$ & \{$\Gamma_{13}$, $\Gamma_{23}$\} & $-$ & $\{\frac{1}{2}[\cos(\mathbf{k}\cdot\boldsymbol{\delta}_{1})-\cos(\mathbf{k}\cdot\boldsymbol{\delta}_{2})], \frac{1}{2\sqrt{3}}[\cos(\mathbf{k}\cdot\boldsymbol{\delta}_{1})+\cos(\mathbf{k}\cdot\boldsymbol{\delta}_{2})-2\cos(\mathbf{k}\cdot\boldsymbol{\delta}_{3})]\}$ & $+$ \\
\hline  \\ [-2ex]
$\tilde{\Gamma}_{3}^{+}$ & \{$\Gamma_{14}$, $\Gamma_{24}$\} & $-$ & $\{-\frac{1}{2\sqrt{3}}[\cos(\mathbf{k}\cdot\boldsymbol{\delta}_{1})+\cos(\mathbf{k}\cdot\boldsymbol{\delta}_{2})-2\cos(\mathbf{k}\cdot\boldsymbol{\delta}_{3})], \frac{1}{2}[\cos(\mathbf{k}\cdot\boldsymbol{\delta}_{1})-\cos(\mathbf{k}\cdot\boldsymbol{\delta}_{2})]\}$ & $+$ \\
\hline  \\ [-2ex]
$\tilde{\Gamma}_{1}^{-}$ & $\Gamma_{3}$ & $-$ & $\frac{1}{3}[\sin(\mathbf{k}\cdot\mathbf{a}_{1})+\sin(\mathbf{k}\cdot\mathbf{a}_{2})+\sin(\mathbf{k}\cdot\mathbf{a}_{3})]$ & $-$ \\
\hline  \\ [-2ex]
$\tilde{\Gamma}_{1}^{-}$ & $\Gamma_{35}$ & $+$ & $\frac{1}{3}[\sin(\mathbf{k}\cdot\mathbf{a}_{1})+\sin(\mathbf{k}\cdot\mathbf{a}_{2})+\sin(\mathbf{k}\cdot\mathbf{a}_{3})]$ & $-$ \\
\hline  \\ [-2ex]
$\tilde{\Gamma}_{2}^{-}$ & $\Gamma_{4}$ & $-$ & $\frac{1}{3}[\sin(\mathbf{k}\cdot\boldsymbol{\delta}_{1})
+\sin(\mathbf{k}\cdot\boldsymbol{\delta}_{2})+\sin(\mathbf{k}\cdot\boldsymbol{\delta}_{3})], \sin(\mathbf{k}\cdot\boldsymbol{\delta}_{4})$ & $-$  \\
\hline  \\ [-2ex]
$\tilde{\Gamma}_{2}^{-}$ & $\Gamma_{45}$ & $+$ & $\frac{1}{3}[\sin(\mathbf{k}\cdot\boldsymbol{\delta}_{1})
+\sin(\mathbf{k}\cdot\boldsymbol{\delta}_{2})+\sin(\mathbf{k}\cdot\boldsymbol{\delta}_{3})], \sin(\mathbf{k}\cdot\boldsymbol{\delta}_{4})$ & $-$  \\
\hline  \\ [-2ex]
$\tilde{\Gamma}_{3}^{-}$ & \{$\Gamma_{1}$, $\Gamma_{2}$\} & $-$ & $\{-\frac{1}{2\sqrt{3}}[\sin(\mathbf{k}\cdot\boldsymbol{\delta}_{1})+\sin(\mathbf{k}\cdot\boldsymbol{\delta}_{2})-2\sin(\mathbf{k}\cdot\boldsymbol{\delta}_{3})], \frac{1}{2}[\sin(\mathbf{k}\cdot\boldsymbol{\delta}_{1})-\sin(\mathbf{k}\cdot\boldsymbol{\delta}_{2})]\}$ & $-$ \\
\hline  \\ [-2ex]
$\tilde{\Gamma}_{3}^{-}$ & \{$\Gamma_{15}$, $\Gamma_{25}$\} & $+$ & $\{-\frac{1}{2\sqrt{3}}[\sin(\mathbf{k}\cdot\boldsymbol{\delta}_{1})+\sin(\mathbf{k}\cdot\boldsymbol{\delta}_{2})-2\sin(\mathbf{k}\cdot\boldsymbol{\delta}_{3})], \frac{1}{2}[\sin(\mathbf{k}\cdot\boldsymbol{\delta}_{1})-\sin(\mathbf{k}\cdot\boldsymbol{\delta}_{2})]\}$ & $-$  \\
\hline
\hline
\end{tabular}
\end{table*}

The $g_{j\alpha}(\mathbf{k},\mathbf{d}_{\nu})$ function is constructed by projection operators pertaining to various irreducible representations.\cite{dresselhaus08} Since only the character table is available, we could only construct the character projection operator which generate symmetrized Fourier functions that are linear combinations of the various components of an irreducible representation. The \emph{character projection operator} is defined as
\begin{equation}
\hat{P}^{(j)}=\frac{n_{j}}{h}\sum\limits_{R}\chi^{(j)}(R)^{\ast}\hat{P}_{R},
\end{equation}
where $j$ labels a certain irreducible representation, $n_{j}$ is the dimension of this representation, $h$ is
total number of symmetry elements $R$ in the group, $\chi^{(j)}(R)$ is the character of $R$ in the $j$-th
irreducible representation, and $\hat{P}_{R}$ is the symmetry operator for the symmetry element $R$. To get
a basis pertaining to wave vector $\mathbf{k}$ and the $j$-th irreducible representation in terms of
$\mathbf{d}_{\nu}$, we act $\hat{P}^{(j)}$ on $e^{i\mathbf{k}\cdot\mathbf{d}_{\nu}}$.
First of all, since $\hat{P}_{\mathcal{C}}$ keeps $e^{i\mathbf{k}\cdot\mathbf{d}_{\nu}}$ invariant, no basis of the prescribed form could be constructed for the six irreducible representations $\tilde{\Gamma}_{4}^{\pm}$, $\tilde{\Gamma}_{5}^{\pm}$, and $\tilde{\Gamma}_{6}^{\pm}$. For $\tilde{\Gamma}_{1}^{+}$, first consider $\mathbf{d}_{0}=(0,0,0)$, we get $\hat{P}^{(\tilde{\Gamma}_{1}^{+})}1=1$. So, a constant could be taken as a basis for (and only for)  $\tilde{\Gamma}_{1}^{+}$. Now consider a NN bond $\boldsymbol{\delta}_{1}=(\frac{\sqrt{3}}{2}a, \frac{1}{2}a, 0)$,
we have $\hat{P}^{(\tilde{\Gamma}_{1}^{+})}e^{i\mathbf{k}\cdot\boldsymbol{\delta}_{1}}=\frac{1}{3}[\cos(\mathbf{k}\cdot\boldsymbol{\delta}_{1})
+\cos(\mathbf{k}\cdot\boldsymbol{\delta}_{2})+\cos(\mathbf{k}\cdot\boldsymbol{\delta}_{3})]$. The result do not change if we replace $\boldsymbol{\delta}_{1}$ by $\boldsymbol{\delta}_{2}$ or $\boldsymbol{\delta}_{3}$. For $\boldsymbol{\delta}_{4}$, we have $\hat{P}^{(\tilde{\Gamma}_{1}^{+})}e^{i\mathbf{k}\cdot\boldsymbol{\delta}_{4}}=\cos(\mathbf{k}\cdot\boldsymbol{\delta}_{4})$. We define the in plane 2NN lattice vectors as $\mathbf{a}_{1}=\boldsymbol{\delta}_{1}-\boldsymbol{\delta}_{2}$,
$\mathbf{a}_{2}=\boldsymbol{\delta}_{2}-\boldsymbol{\delta}_{3}$, and $\mathbf{a}_{3}=\boldsymbol{\delta}_{3}-\boldsymbol{\delta}_{1}$. A basis for $\tilde{\Gamma}_{1}^{+}$ could also be constructed in terms of these 2NN bonds, which turns out to be $\frac{1}{3}[\cos(\mathbf{k}\cdot\mathbf{a}_{1})
+\cos(\mathbf{k}\cdot\mathbf{a}_{2})+\cos(\mathbf{k}\cdot\mathbf{a}_{3})]$. However, we restrict to NN bonds if it could be used to construct a basis set, to keep the model minimal.

Basis functions for other representations up to 2NN in plane bonds are similarly constructed. For $\tilde{\Gamma}_{1}^{-}$ representation, a direct calculation shows that $\hat{P}^{(\tilde{\Gamma}_{1}^{-})}e^{i\mathbf{k}\cdot\boldsymbol{\delta}_{l}}=0$ for $l=1,2,3,4$. So no basis could be constructed in terms of the NN bonds. Consider the 2NN bond $\mathbf{a}_{1}$, we have $\hat{P}^{(\tilde{\Gamma}_{1}^{-})}e^{i\mathbf{k}\cdot\mathbf{a}_{1}}=\frac{i}{3}[\sin(\mathbf{k}\cdot\mathbf{a}_{1})
+\sin(\mathbf{k}\cdot\mathbf{a}_{2})+\sin(\mathbf{k}\cdot\mathbf{a}_{3})]$. For $\tilde{\Gamma}_{2}^{+}$ representation, calculation shows that no basis could be constructed up to 2NN bonds. But since the $\tilde{\Gamma}_{2}^{+}$ representation does not appear in the $\mathbf{k}$$\cdot$$\mathbf{p}$ model\cite{liu10}, we would still restrict our model within 2NN bonds. For $\tilde{\Gamma}_{2}^{-}$ representation, we get $\hat{P}^{(\tilde{\Gamma}_{2}^{-})}e^{i\mathbf{k}\cdot\boldsymbol{\delta}_{1}}=\frac{i}{3}[\sin(\mathbf{k}\cdot\boldsymbol{\delta}_{1})
+\sin(\mathbf{k}\cdot\boldsymbol{\delta}_{2})+\sin(\mathbf{k}\cdot\boldsymbol{\delta}_{3})]$, $\hat{P}^{(\tilde{\Gamma}_{2}^{-})}e^{i\mathbf{k}\cdot\boldsymbol{\delta}_{4}}=i\sin(\mathbf{k}\cdot\boldsymbol{\delta_{4}})$, and
$\hat{P}^{(\tilde{\Gamma}_{2}^{-})}e^{i\mathbf{k}\cdot\mathbf{a}_{l}}=0$ ($l=1,2,3$).

$\tilde{\Gamma}_{3}^{\pm}$ are two dimensional representations. For them, the character projection operators would in general generate a linear combination of two basis functions when operating it on an arbitrary Fourier exponential. If an arbitrary set of basis functions are required, we could take this as one basis and generate another basis which is orthogonal to it to form a basis set. However, since we would form invariants in terms of these symmetrized Fourier functions and the basis matrix functions, the two sets of basis functions should transform identically under the group operation. Enforcing this requirement, we could get the proper sets of basis Fourier functions which have the same group transformation properties as the corresponding basis matrix functions. The symmetrized Fourier functions are thus as shown in the fourth column of Table II. The time reversal property of the basis Fourier functions are as shown in the fifth column under the title of $T'$. In the fourth column of Table II, two functions in a single brace form a basis set for the corresponding two dimensional representation. The non-braced functions are optional bases for the corresponding one dimensional representation.

Having the basis matrix functions and the symmetrized Fourier functions at hand, the tight binding model is constructed by multiplying the corresponding components of the two together to form invariants of the symmetry group. Since the material and hence the model for it preserves time reversal symmetry, the terms to be multiplied together to form invariants should also have the same $T$ value. We are thus lead by the requirement of group symmetry, time reversal invariance, and Hermiticity to write $H_{0}(\mathbf{k})$ in the most general form (with the restriction of keeping only short-range hoppings) as
\begin{eqnarray}
H_{0}(\mathbf{k})&=&\epsilon(\mathbf{k})I_{4}+M(\mathbf{k})\Gamma_5+B_{0}c_{z}(\mathbf{k})\Gamma_{4}
+A_{0}[c_{y}(\mathbf{k})\Gamma_{1}   \notag \\
&&-c_{x}(\mathbf{k})\Gamma_{2}]+R_{1}d_{1}(\mathbf{k})\Gamma_{3}+R_{2}d_{2}(\mathbf{k})\Gamma_{4}.
\end{eqnarray}
Definitions of the various terms are the same as in the main text.

To determine the parameters in the model, we compare the model with the $\mathbf{k}\cdot\mathbf{p}$ model defined close to $\boldsymbol{\Gamma}=(0,0,0)$. So, we expand the various terms close to $\boldsymbol{\Gamma}$ as $1-\cos(\mathbf{k}\cdot\boldsymbol{\delta}_{4})\simeq\frac{1}{2}k_{z}^{2}c^{2}$, $3-\cos(\mathbf{k}\cdot\boldsymbol{\delta}_{1})-\cos(\mathbf{k}\cdot\boldsymbol{\delta}_{2})-\cos(\mathbf{k}\cdot\boldsymbol{\delta}_{3})\simeq\frac{1}{2}(k_{x}^{2}+k_{y}^{2})a^{2}$,
$\sin(\mathbf{k}\cdot\boldsymbol{\delta}_{4})\simeq k_{z}c$, $\frac{1}{3}[\sin(\mathbf{k}\cdot\boldsymbol{\delta}_{1})+\sin(\mathbf{k}\cdot\boldsymbol{\delta}_{2})-2\sin(\mathbf{k}\cdot\boldsymbol{\delta}_{3})]
\simeq k_{y}a$, $\frac{1}{\sqrt{3}}[\sin(\mathbf{k}\cdot\boldsymbol{\delta}_{1})-\sin(\mathbf{k}\cdot\boldsymbol{\delta}_{2})]\simeq k_{x}a$, $\sin(\mathbf{k}\cdot\mathbf{a}_{1})+\sin(\mathbf{k}\cdot\mathbf{a}_{2})+\sin(\mathbf{k}\cdot\mathbf{a}_{3})\simeq \frac{3\sqrt{3}}{8}(3k_{x}k_{y}^{2}-k_{x}^{3})a^{3}$, and $\sin(\mathbf{k}\cdot\boldsymbol{\delta}_{1})+\sin(\mathbf{k}\cdot\boldsymbol{\delta}_{2})+\sin(\mathbf{k}\cdot\boldsymbol{\delta}_{3})
\simeq\frac{1}{8}(k_{y}^{3}-3k_{x}^{2}k_{y})a^{3}$. Substituting these approximations into the above model, it clearly has a form identical to the $\mathbf{k}$$\cdot$$\mathbf{p}$ model proposed by Liu \emph{et al}.\cite{liu10} Demanding that our tight binding model reduce to the same model as that used by Liu \emph{et al}, we could derive the values of the various parameters as shown in Table I of the main text. A calculation of the bulk density of states shows that the bulk energy gaps of Bi$_{2}$Se$_{3}$ and Bi$_{2}$Te$_{3}$ corresponding to the above parameters are both too small to be comparable to experiment. Test calculations show that changing $M_{1}$ to 0.62 eV for Bi$_2$Se$_3$ and 0.102 eV for Bi$_2$Te$_3$ and keeping other parameters unchanged, the bulk energy gaps of Bi$_2$Se$_3$ and  Bi$_2$Te$_3$ are approximately 0.26 eV and 0.06 eV, which are close to known experimental and first principle theoretical results.\cite{zhang09}

\section{\label{sec:exchange}Superexchange Coupling Terms}

We derive the dominant superexchange couplings that could arise from the model proposed in the main text. The model combines the tight binding model and the short range correlation terms and is written as $H=H_{0}+H_{1}$, with
\begin{equation}
H_{1}=U\sum\limits_{\mathbf{i}\alpha}\hat{n}_{\mathbf{i}\alpha\uparrow}\hat{n}_{\mathbf{i}\alpha\downarrow}
+V\sum\limits_{\mathbf{i}}\hat{n}_{\mathbf{i}}^{a}\hat{n}_{\mathbf{i}}^{b},
\end{equation}
in which $\alpha$ runs over the two orbitals $a$ and $b$. While the actual strength of the on-site correlations for Bi$_{2}$X$_{3}$ (X is Se or Te) materials are presently unknown, evidences have appeared that they should be sizeable to explain the experimental findings.\cite{craco12,giraud11} In light of this information, we could consider a strongly correlated system when deriving the low energy effective model. Working with the two orbital model $H_{0}$, the materials are doped semiconductors close to half filling. We thus generalize the routine procedure of deriving the $t-J$ model from the one orbital Hubbard model to the present two-orbital model \cite{schrieffer66}, first get the effective model at half filling and then dope the model to approximately represent the actual materials.

Since the two orbitals within a single unit cell reside on different sites, we expect that $U>V>0$ holds. At half filling and for sufficiently large $U$ and $V$, the system would in most of the time be restricted within the subspace of local intraorbital single occupation. Projecting out the subspace with doubly occupied orbitals, we have\cite{schrieffer66}
\begin{equation}
\tilde{H}=PHP-\frac{1}{U}PHQHP,
\end{equation}
where $P$ and $Q$ are projection operators which project into the subspace of intraorbital no-double-occupancy and the subspace with intraorbital double occupancy (and thus with empty orbital in the case of half filling of interest), respectively. $P^2=P$ and $Q^2=Q=1-P$. In the present two orbital systems, virtual hoppings could be intraorbital or interorbital, the excitation energy of which are both $U$. The second term in $\tilde{H}$ is the exchange term arising from mixing of the two subspaces and would be denoted as $H_{ex}$. Since $H_{0}$ has many terms, we expect that a lot of different terms would appear for $H_{ex}$ when the multiplication is carried out. However, we note that the parameters in Table I of the main text show very large differences. The different superexchange terms thus derived would also show very large differences. Since we are interested in finding out the dominant pairing channel, it is sufficient to retain only the largest several superexchange terms mediated by the hopping terms corresponding to the largest several parameters in Table I (the main text). We thus retain the superexchange terms mediated by $C_2$, $M_2$ and $R_2$, which are apparently larger than other parameters in Table I (the main text). As a further approximation, we retain only two-site terms. Since before the virtual hop or after two complementary virtual hops the system only has singly occupied orbitals, the two complementary hop must both be intra-orbital or inter-orbital.

Straightforward deductions show that $C_{2}$ mediates an intraorbital superexchange
\begin{equation}
H_{intra}=\frac{8C_{2}^{2}}{9U}\sum\limits_{\mathbf{i}\boldsymbol{\delta}\alpha}
(\mathbf{S}_{\mathbf{i}\alpha}\cdot\mathbf{S}_{\mathbf{i}+\boldsymbol{\delta},\alpha}
-\frac{1}{4}\hat{n}_{\mathbf{i}}^{\alpha}\hat{n}_{\mathbf{i}+\boldsymbol{\delta}}^{\alpha}),
\end{equation}
where $\mathbf{i}$ runs over unit cells, $\boldsymbol{\delta}$ runs over the six NN in plane bonds $\pm\boldsymbol{\delta}_{j}$ ($j$=1, 2, 3), and the $\alpha$ summation runs over the two orbitals.
$\hat{n}_{\mathbf{i}}^{\alpha}=\hat{n}_{\mathbf{i}\alpha\uparrow}+\hat{n}_{\mathbf{i}\alpha\downarrow}$
is the electron number operator for $\alpha$ orbital. Written as above, the single occupation condition per unit cell and per orbital is imposed. That is, we use $\hat{n}_{\mathbf{i}\alpha\uparrow}+\hat{n}_{\mathbf{i}\alpha\downarrow}=1$ to simplify terms. For example, $(1-\hat{n}_{\mathbf{i}\alpha\bar{\sigma}})a_{\mathbf{i}\sigma}^{\dagger}\simeq \hat{n}_{\mathbf{i}\alpha\sigma}a_{\mathbf{i}\sigma}^{\dagger}=a_{\mathbf{i}\sigma}^{\dagger}$ ($\bar{\sigma}$ is the opposite spin of $\sigma$).

Following the same convention, the superexchange terms mediated by $M_2$ and $R_2$ are obtained
\begin{equation}
H_{M_{2}}=\frac{(4M_{2})^2}{9U}\sum\limits_{\mathbf{i}\boldsymbol{\delta}}
(\mathbf{S}_{\mathbf{i}a}\cdot\mathbf{S}_{\mathbf{i}+\boldsymbol{\delta},b}
-\frac{1}{4}\hat{n}_{\mathbf{i}}^{a}\hat{n}_{\mathbf{i}+\boldsymbol{\delta}}^{b}),
\end{equation}
\begin{equation}
H_{R_{2}}=\frac{(8R_{2})^2}{U}\sum\limits_{\mathbf{i}\boldsymbol{\delta}}
(\mathbf{S}_{\mathbf{i}a}\cdot\mathbf{S}_{\mathbf{i}+\boldsymbol{\delta},b}
-\frac{1}{4}\hat{n}_{\mathbf{i}}^{a}\hat{n}_{\mathbf{i}+\boldsymbol{\delta}}^{b}).
\end{equation}
Besides the the above two terms, the crossing terms between $M_2$ and $R_2$ also mediate a superexchange term in the same interorbital channel, which turns out to be
\begin{equation}
H_{M_2R_2}=\frac{64M_{2}R_{2}}{3U}\sum\limits_{\mathbf{i}\boldsymbol{\delta}}
(-1)^{\eta(\boldsymbol{\delta})}(\mathbf{S}_{\mathbf{i}a}\cdot\mathbf{S}_{\mathbf{i}+\boldsymbol{\delta},b}
-\frac{1}{4}\hat{n}_{\mathbf{i}}^{a}\hat{n}_{\mathbf{i}+\boldsymbol{\delta}}^{b}).
\end{equation}
$\eta(\boldsymbol{\delta})$ is defined as $0$ ($1$) for $\boldsymbol{\delta}=\boldsymbol{\delta}_{j}$
($\boldsymbol{\delta}=-\boldsymbol{\delta}_{j}$), with $j$=1, 2, 3. Arising from the crossing term between $M_{2}$ and $R_{2}$, $\eta(\boldsymbol{\delta})$ inherits the bond-wise sign change character from $R_{2}$. Whereas the sign change in $R_{2}$ hopping is canceled out in the $R_{2}^{2}$ term. It is clear that the coefficients of the above three terms combine nicely into a square term, written as $J^{ab}_{\mathbf{i},\mathbf{i}+\boldsymbol{\delta}}=
\frac{(8R_{2})^2}{U}[1+(-1)^{\eta(\boldsymbol{\delta})}\frac{M_{2}}{6R_{2}}]^{2}$. Thus the interorbital superexchange term is written as
\begin{equation}
H_{inter}=\sum\limits_{\mathbf{i}\boldsymbol{\delta}}
J^{ab}_{\mathbf{i},\mathbf{i}+\boldsymbol{\delta}}
(\mathbf{S}_{\mathbf{i}a}\cdot\mathbf{S}_{\mathbf{i}+\boldsymbol{\delta},b}
-\frac{1}{4}\hat{n}_{\mathbf{i}}^{a}\hat{n}_{\mathbf{i}+\boldsymbol{\delta}}^{b}).
\end{equation}
Since the coupling constants in both $H_{intra}$ and $H_{inter}$ are positive, the above superexchange terms favor only spin singlet pairings. The other hopping terms in $H_{0}$, such as the $A_{0}$ term, might favor triplet pairings. A similar derivation shows that it is indeed the case. However, since it is much smaller than the above two terms, the pairing favored by it can not survive the competition. On the other hand, though the spin-orbit interaction in $H_{0}$ mixes spin singlet and spin triplet components, the induced spin triplet component is always much smaller than the spin singlet component.

Finally, we mention that the interorbital constant term in $H_{0}(\mathbf{k})$, that is $(M_{0}+2M_{1}+4M_{2})\Gamma_{5}$, also mediates an intra-unit-cell interorbital superexchange term. This term is not negligible compared to the terms retained in the main text. However, this term favors a $\mathbf{k}$-independent interorbital spin singlet pairing, which is equivalent to the third or fourth pairing defined in the main text by setting $\varphi_{3(4)}(\mathbf{k})=1$ and was already studied in previous works \cite{fu10,hao11}. So, we have ignored this superexchange term in this study which focuses on anisotropic spin singlet pairings. But, as mentioned in the main text, this fully gapped interorbital spin singlet pairing component could readily be incorporated into our framework to account for some experimental features, such as the specific heat experiments which indicates the presence of a fully gapped pairing component.\cite{kriener11}

\section{\label{sec:MFA} Pairing Symmetry and Mean Field Calculations}

With the above model at hand, the simplest and fastest way to evaluate the most probable pairing symmetry is to perform a mean field calculation.\cite{seo08,kotliar88,goswami10} Here, we show some details of the mean field calculations and analysis of the results. In order to derive proper superexchange type of terms that favor pairing, we have imposed a condition of strong correlation. However, no firm evidences showing that the system belongs to strong correlation \emph{limit} have appeared. So, we do not impose the projection of $P$ on $H_{0}$. Instead, we keep the on-site correlation terms together with $H_{0}$ unchanged to qualitatively reflect the constraint on intraorbital double occupancy. Thus, we use a $t-U-V-J$ model, that is $H=H_{0}+H_{1}+H_{ex}$, as a starting point to perform the mean field analyses.

At the mean filed level, pairing comes only from $H_{ex}$. In a full mean field decoupling over $H_{ex}$, we would get not only pairing terms but also terms for magnetic orders and various bond orders. However, since on one hand we are here interested only in the superconducting phase and on the other hand no evidences for the existence of other phases have appeared, we would keep only the decoupling channel that leads to pairing. In terms of the identity $\boldsymbol{\sigma}_{\alpha\beta}\cdot\boldsymbol{\sigma}_{\alpha'\beta'}=2\delta_{\alpha\beta'}\delta_{\alpha'\beta}
-\delta_{\alpha\beta}\delta_{\alpha'\beta'}$ for the scalar product of two Pauli vectors, we get
\begin{eqnarray}
&&\mathbf{S}_{\mathbf{i}a}\cdot\mathbf{S}_{\mathbf{i}+\boldsymbol{\delta},a}
-\frac{1}{4}\hat{n}_{\mathbf{i}}^{a}\hat{n}_{\mathbf{i}+\boldsymbol{\delta}}^{a}  \notag \\
&&=-\frac{1}{2}(a_{\mathbf{i}\uparrow}^{\dagger}a_{\mathbf{i}+\boldsymbol{\delta},\downarrow}^{\dagger}
-a_{\mathbf{i}\downarrow}^{\dagger}a_{\mathbf{i}+\boldsymbol{\delta},\uparrow}^{\dagger})
(a_{\mathbf{i}+\boldsymbol{\delta},\downarrow}a_{\mathbf{i}\uparrow}
-a_{\mathbf{i}+\boldsymbol{\delta},\uparrow}a_{\mathbf{i}\downarrow}),  \notag \\
&&\mathbf{S}_{\mathbf{i}b}\cdot\mathbf{S}_{\mathbf{i}+\boldsymbol{\delta},b}
-\frac{1}{4}\hat{n}_{\mathbf{i}}^{b}\hat{n}_{\mathbf{i}+\boldsymbol{\delta}}^{b}  \notag \\
&&=-\frac{1}{2}(b_{\mathbf{i}\uparrow}^{\dagger}b_{\mathbf{i}+\boldsymbol{\delta},\downarrow}^{\dagger}
-b_{\mathbf{i}\downarrow}^{\dagger}b_{\mathbf{i}+\boldsymbol{\delta},\uparrow}^{\dagger})
(b_{\mathbf{i}+\boldsymbol{\delta},\downarrow}b_{\mathbf{i}\uparrow}
-b_{\mathbf{i}+\boldsymbol{\delta},\uparrow}b_{\mathbf{i}\downarrow}),  \notag \\
&&\mathbf{S}_{\mathbf{i}a}\cdot\mathbf{S}_{\mathbf{i}+\boldsymbol{\delta},b}
-\frac{1}{4}\hat{n}_{\mathbf{i}}^{a}\hat{n}_{\mathbf{i}+\boldsymbol{\delta}}^{b}  \notag \\
&&=-\frac{1}{2}(a_{\mathbf{i}\uparrow}^{\dagger}b_{\mathbf{i}+\boldsymbol{\delta},\downarrow}^{\dagger}
-a_{\mathbf{i}\downarrow}^{\dagger}b_{\mathbf{i}+\boldsymbol{\delta},\uparrow}^{\dagger})
(b_{\mathbf{i}+\boldsymbol{\delta},\downarrow}a_{\mathbf{i}\uparrow}
-b_{\mathbf{i}+\boldsymbol{\delta},\uparrow}a_{\mathbf{i}\downarrow}), \notag \\
\end{eqnarray}
where $\boldsymbol{\delta}=\pm\boldsymbol{\delta}_{1}$, $\pm\boldsymbol{\delta}_{2}$, and $\pm\boldsymbol{\delta}_{3}$. Since the coefficients for the intraorbital and interorbital superexchange couplings are both positive, the above decomposition makes it clear that $H_{ex}$ favors and only favors spin singlet pairings. Take advantage of the above expressions, eighteen independent mean field parameters are introduced to describe the superconducting pairing. Firstly, we define twelve operators $\hat{\chi}_{a}^{\nu\pm}=
a_{\mathbf{i}\pm\boldsymbol{\delta}_{\nu},\downarrow}a_{\mathbf{i}\uparrow}
-a_{\mathbf{i}\pm\boldsymbol{\delta}_{\nu},\uparrow}a_{\mathbf{i}\downarrow}$
($\nu=1, 2, 3$) and $\hat{\chi}_{b}^{\nu\pm}=
b_{\mathbf{i}\pm\boldsymbol{\delta}_{\nu},\downarrow}b_{\mathbf{i}\uparrow}
-b_{\mathbf{i}\pm\boldsymbol{\delta}_{\nu},\uparrow}b_{\mathbf{i}\downarrow}$
($\nu=1, 2, 3$). Their expectation values, $\chi_{a}^{\nu\pm}=\langle\hat{\chi}_{a}^{\nu\pm}\rangle$ and $\chi_{b}^{\nu\pm}=\langle\hat{\chi}_{b}^{\nu\pm}\rangle$, define the twelve mean field parameters for the intraorbital spin singlet pairing. Then we define another six operators $\hat{\chi}_{ba}^{\nu\pm}=
b_{\mathbf{i}\pm\boldsymbol{\delta}_{\nu},\downarrow}a_{\mathbf{i}\uparrow}
-b_{\mathbf{i}\pm\boldsymbol{\delta}_{\nu},\uparrow}a_{\mathbf{i}\downarrow}$
($\nu=1, 2, 3$). Their expectation values, $\chi_{ba}^{\nu\pm}=\langle\hat{\chi}_{ba}^{\nu\pm}\rangle$, define the six mean field parameters for the interorbital spin singlet pairing.

Retaining only the decoupling to the superconducting channel, we make the mean field approximation to $H_{ex}$ in terms of $(\hat{\chi}_{\alpha}^{\nu\pm})^{\dagger}\hat{\chi}_{\alpha}^{\nu\pm}\simeq(\chi_{\alpha}^{\nu\pm})^{\ast}\hat{\chi}_{\alpha}^{\nu\pm}
+(\hat{\chi}_{\alpha}^{\nu\pm})^{\dagger}\chi_{\alpha}^{\nu\pm}-(\chi_{\alpha}^{\nu\pm})^{\ast}\chi_{\alpha}^{\nu\pm}$ ($\alpha$ is $a$ or $b$), and $(\hat{\chi}_{ba}^{\nu\pm})^{\dagger}\hat{\chi}_{ba}^{\nu\pm}\simeq(\chi_{ba}^{\nu\pm})^{\ast}\hat{\chi}_{ba}^{\nu\pm}
+(\hat{\chi}_{ba}^{\nu\pm})^{\dagger}\chi_{ba}^{\nu\pm}-(\chi_{ba}^{\nu\pm})^{\ast}\chi_{ba}^{\nu\pm}$.\cite{seo08,kotliar88,goswami10} For $H_{1}$, it is easy to see that it does not favor superconducting state at the mean field level. We make the mean field decoupling to $H_{1}$ in the simplest way as $\hat{n}_{1}\hat{n}_{2}\rightarrow\langle\hat{n}_{1}\rangle\hat{n}_{2}+\hat{n}_{1}\langle\hat{n}_{2}\rangle
-\langle\hat{n}_{1}\rangle\langle\hat{n}_{2}\rangle$, which introduces four mean field parameters $n_{\alpha\sigma}$ ($\alpha$ is for the $a$ or $b$ orbital, $\sigma$ is for the $\uparrow$ or $\downarrow$ spin).

After the above mean field decoupling, the Hamiltonian is now a bilinear of electron operators and is easily transformed into the reciprocal space. Then the mean field calculation is performed in a self-consistent manner starting from an arbitrary set of initial values for the $18$ spin singlet pairing amplitudes and $4$ on-site occupation numbers. The mean field calculation turns out to converge very well. When convergence is arrived at for a certain set of parameters, we analyze the pairings that are contained in the results.

For the spin singlet solution obtained by the above self-consistent mean field calculation, we are interested in those pairing components contained in it which hold symmetry compatible with the crystal symmetry. In addition, we focus on time reversal invariant pairings. At first sight, the spin singlet pairing terms in the superconducting Hamiltonian are to be constructed completely parallel to the construction of $H_{0}(\mathbf{k})$. However, since the 4$\times$4 pairing term which appears in the off-diagonal position of the Bogoliubov-de Gennes (BdG) Hamiltonian needs \emph{not} be Hermitian, the number of possible time reversal invariant combinations for it is increased. A direct survey over Table II shows that the four one dimensional representations $\tilde{\Gamma}_{1}^{\pm}$ and $\tilde{\Gamma}_{2}^{\pm}$ all describes spin triplet pairings, so drop out of our present analysis. For the remaining two dimensional representations $\tilde{\Gamma}_{3}^{\pm}$, it is interesting that for each of the four realizations of them, one basis matrix function corresponds to spin singlet pairing while the other basis matrix function corresponds to spin triplet pairing. For example, for the $\{\Gamma_{13},\Gamma_{23}\}$ realization of $\tilde{\Gamma}_{3}^{+}$, $\Gamma_{13}=-s_{2}\otimes\sigma_{0}$ is in the spin singlet channel while $\Gamma_{23}=s_{1}\otimes\sigma_{0}$ describes spin triplet pairing. In addition, $\Gamma_{24}=s_{2}\otimes\sigma_{1}$ for $\tilde{\Gamma}_{3}^{+}$, $\Gamma_{2}=s_{2}\otimes\sigma_{3}$ and $\Gamma_{25}=s_{2}\otimes\sigma_{2}$ for $\tilde{\Gamma}_{3}^{-}$ all belong to the spin singlet channel.

The coexistence of spin singlet and spin triplet pairing components in a single irreducible representation might be a direct result of the presence of spin-orbit interaction, which makes the pairing with a definite spin state not well defined. That is, even though the correlation term favors only spin singlet pairing, some spin triplet pairing component would be induced from the spin singlet pairing by the spin-orbit interaction. In the present study, we do not analyze the spin triplet pairings induced by the spin-orbit interaction and concentrate on the dominant spin singlet pairing components. So, we have four basis matrices pertaining to two irreducible representations that are possibly of interest. The wave vector dependence of the pairings are taken from the basis Fourier functions in Table II. Since now only one component of the basis matrix functions exists for a certain set of the two dimensional representation, we consider both of the two basis Fourier functions as possible candidates, since a symmetry transformation would mix the two basis Fourier functions and the absence of spin triplet pairing then leaves a product of a linear combination of the two basis Fourier functions with the spin singlet basis matrix function.

In the above convention, we would get eight independent spin singlet pairings. The symmetry factors ($\mathbf{k}$-dependency) of the pairings are defined as $\varphi_{1}(\mathbf{k})=\cos(\mathbf{k}\cdot\boldsymbol{\delta}_{1})-\cos(\mathbf{k}\cdot\boldsymbol{\delta}_{2})$
and
$\varphi_{2}(\mathbf{k})=2\cos(\mathbf{k}\cdot\boldsymbol{\delta}_{3})-\cos(\mathbf{k}\cdot\boldsymbol{\delta}_{1})
-\cos(\mathbf{k}\cdot\boldsymbol{\delta}_{2})$, $\varphi_{3(4)}(\mathbf{k})=\varphi_{1(2)}(\mathbf{k})$, $\varphi_{5}(\mathbf{k})=\sin(\mathbf{k}\cdot\boldsymbol{\delta}_{1})-\sin(\mathbf{k}\cdot\boldsymbol{\delta}_{2})$
and
$\varphi_{6}(\mathbf{k})=\sin(\mathbf{k}\cdot\boldsymbol{\delta}_{1})+\sin(\mathbf{k}\cdot\boldsymbol{\delta}_{2})
-2\sin(\mathbf{k}\cdot\boldsymbol{\delta}_{3})$, $\varphi_{7(8)}(\mathbf{k})=\varphi_{5(6)}(\mathbf{k})$.
The first two pairings correspond to $\Gamma_{31}=-\Gamma_{13}$ and are $\Delta_{j}\phi_{\mathbf{k}}^{\dagger}i\Gamma_{31}(\phi_{-\mathbf{k}}^{\dagger})^{\text{T}}\varphi_{j}(\mathbf{k})$
($j$=1, 2). The next two pairings correspond to $\Gamma_{24}$ and are $\Delta_{j}\phi_{\mathbf{k}}^{\dagger}i\Gamma_{24}(\phi_{-\mathbf{k}}^{\dagger})^{\text{T}}\varphi_{j}(\mathbf{k})$
($j$=3, 4). The fifth and sixth pairings corresponds to $\Gamma_{25}$ and are $\Delta_{j}\phi_{\mathbf{k}}^{\dagger}i\Gamma_{25}(\phi_{-\mathbf{k}}^{\dagger})^{\text{T}}\varphi_{j}(\mathbf{k})$
($j$=5, 6). The last two pairings correspond to $\Gamma_{2}$, and are written as $\Delta_{j}\phi_{\mathbf{k}}^{\dagger}\Gamma_{2}(\phi_{-\mathbf{k}}^{\dagger})^{\text{T}}\varphi_{j}(\mathbf{k})$
($j$=7, 8). $\Delta_{j}$ ($j=1,\ldots,8$) are the amplitudes of the corresponding pairing components. However, since the summation of the last two pairings over wave vectors in the BZ vanishes, that is $\sum_{\mathbf{k}}\phi_{\mathbf{k}}^{\dagger}\Gamma_{2}(\phi_{-\mathbf{k}}^{\dagger})^{\text{T}}\varphi_{j}(\mathbf{k})=0$ ($j$=7, 8), we only have six spin singlet time reversal invariant parings that are compatible with crystal symmetry and are anisotropic.

For a set of parameters such as a specific $U$ and $V$ and doping level, after obtaining the $18$ pairing order parameters ($\chi_{a}^{\nu\pm}$, $\chi_{b}^{\nu\pm}$, and $\chi_{ba}^{\nu\pm}$, $\nu$=$1$, $2$, $3$) through self-consistent calculations, we could extract the amplitudes for the six symmetry channels defined above.\cite{seo08,kotliar88,goswami10} That is, we could express the value of $\Delta_{i}$ ($i$=$1$, $\ldots$, $6$) in terms of the $18$ mean field pairing order parameters. Since the six pairings are mutually independent, we could get the representation of their amplitudes in terms of all the self-consistent pairing fields. This is as much as to say, when extracting the pairing amplitude of a certain channel among the six possibilities, we regard it as the only pairing that is contained in the self-consistent solution. First consider $\Delta_{1}$. The upper-right $4\times4$ block of the BdG Hamiltonian for this pairing is written explicitly as
\begin{eqnarray*}
&&\Delta_{1}\sum\limits_{\mathbf{k}}\phi^{\dagger}_{\mathbf{k}}i\Gamma_{13}(\phi^{\dagger}_{-\mathbf{k}})^{\text{T}}\varphi_{1}(\mathbf{k})  \notag \\
&&=\Delta_{1}\sum\limits_{\mathbf{k}}[-a^{\dagger}_{\mathbf{k}\uparrow}a^{\dagger}_{-\mathbf{k}\downarrow}+a^{\dagger}_{\mathbf{k}\downarrow}a^{\dagger}_{-\mathbf{k}\uparrow}
-b^{\dagger}_{\mathbf{k}\uparrow}b^{\dagger}_{-\mathbf{k}\downarrow}+b^{\dagger}_{\mathbf{k}\downarrow}b^{\dagger}_{-\mathbf{k}\uparrow}]   \notag  \\
&& \cdot[\cos(\mathbf{k}\cdot\boldsymbol{\delta}_{1})-\cos(\mathbf{k}\cdot\boldsymbol{\delta}_{2})]    \notag  \\
&&=\frac{1}{2}\Delta_{1}\sum\limits_{\mathbf{i}}[(a^{\dagger}_{\mathbf{i}\uparrow}a^{\dagger}_{\mathbf{i}+\boldsymbol{\delta}_{2},\downarrow}
-a^{\dagger}_{\mathbf{i}\downarrow}a^{\dagger}_{\mathbf{i}+\boldsymbol{\delta}_{2},\uparrow}
+a^{\dagger}_{\mathbf{i}\uparrow}a^{\dagger}_{\mathbf{i}-\boldsymbol{\delta}_{2},\downarrow}  \notag \\
&&-a^{\dagger}_{\mathbf{i}\downarrow}a^{\dagger}_{\mathbf{i}-\boldsymbol{\delta}_{2},\uparrow}
-a^{\dagger}_{\mathbf{i}\uparrow}a^{\dagger}_{\mathbf{i}+\boldsymbol{\delta}_{1},\downarrow}
+a^{\dagger}_{\mathbf{i}\downarrow}a^{\dagger}_{\mathbf{i}+\boldsymbol{\delta}_{1},\uparrow}   \notag  \\
&&-a^{\dagger}_{\mathbf{i}\uparrow}a^{\dagger}_{\mathbf{i}-\boldsymbol{\delta}_{1},\downarrow}
+a^{\dagger}_{\mathbf{i}\downarrow}a^{\dagger}_{\mathbf{i}-\boldsymbol{\delta}_{1},\uparrow})+(a\rightarrow b)].
\end{eqnarray*}
The above pairing term is then compared with the mean field decoupling to $H_{ex}$. Requiring the identical terms to be equal to each other, we obtain the representation of $\Delta_{1}$ in terms of the mean field pairing amplitudes as
\begin{eqnarray}
\Delta_{1}&=&\frac{1}{8}J_{1}[\chi_{a}^{1+}+\chi_{a}^{1-}-\chi_{a}^{2+}-\chi_{a}^{2-}  \notag  \\
&&+\chi_{b}^{1+}+\chi_{b}^{1-}-\chi_{b}^{2+}-\chi_{b}^{2-}],
\end{eqnarray}
where $J_{1}=\frac{8C_{2}^{2}}{9U}$. Furthermore, since we consider a uniform solution, the pairing fields are independent of lattice sites. So the above value for $\Delta_{1}$ could be expressed in the wave vector representation as
\begin{equation}
\Delta_{1}=\frac{J_{1}}{2N}\sum\limits_{\mathbf{k}}d(\mathbf{k})\varphi_{1}(\mathbf{k}),
\end{equation}
where $N$ is number of unit cells in the lattice, $d(\mathbf{k})=\langle a_{-\mathbf{k}\downarrow}a_{\mathbf{k}\uparrow}+b_{-\mathbf{k}\downarrow}b_{\mathbf{k}\uparrow}\rangle$.

Amplitudes for the other five pairings could similarly be extracted from a solution of the $18$ mean field pairings. They are written in the wave vector representation as follows.
\begin{equation}
\Delta_{2}=\frac{J_{1}}{6N}\sum\limits_{\mathbf{k}}d(\mathbf{k})\varphi_{2}^{'}(\mathbf{k}),
\end{equation}
where $\varphi_{2}^{'}(\mathbf{k})=\cos(\mathbf{k}\cdot\boldsymbol{\delta}_{3})-2\cos(\mathbf{k}\cdot\boldsymbol{\delta}_{1})
-2\cos(\mathbf{k}\cdot\boldsymbol{\delta}_{2})$.
\begin{equation}
\Delta_{3}=-\frac{(2M_{2})^2}{9NU}\sum\limits_{\mathbf{k}}d^{'}(\mathbf{k})[(1+\frac{36R_{2}^{2}}{M^{2}})\varphi_{3}(\mathbf{k})
-i\frac{12R_{2}}{M_{2}}\varphi_{5}(\mathbf{k})],
\end{equation}
where $d^{'}(\mathbf{k})=\langle b_{-\mathbf{k}\downarrow}a_{\mathbf{k}\uparrow}-b_{-\mathbf{k}\uparrow}a_{\mathbf{k}\downarrow}\rangle$.
\begin{equation}
\Delta_{4}=-\frac{(2M_{2})^2}{27NU}\sum\limits_{\mathbf{k}}d^{'}(\mathbf{k})[(1+\frac{36R_{2}^{2}}{M_{2}^{2}})\varphi_{4}^{'}(\mathbf{k})
+i\frac{24R_{2}}{M_{2}}\varphi_{6}^{'}(\mathbf{k})],
\end{equation}
where $\varphi_{4}^{'}(\mathbf{k})=\varphi_{2}^{'}(\mathbf{k})$ and $\varphi_{6}^{'}(\mathbf{k})=2\sin(\mathbf{k}\cdot\boldsymbol{\delta}_{1})+2\sin(\mathbf{k}\cdot\boldsymbol{\delta}_{2})
-\sin(\mathbf{k}\cdot\boldsymbol{\delta}_{3})$.
\begin{equation}
\Delta_{5}=-\frac{(2M_{2})^2}{9NU}\sum\limits_{\mathbf{k}}d^{'}(\mathbf{k})[i(1+\frac{36R_{2}^{2}}{M_{2}^{2}})\varphi_{5}(\mathbf{k})
-\frac{12R_{2}}{M_{2}}\varphi_{3}(\mathbf{k})].
\end{equation}
\begin{equation}
\Delta_{6}=-\frac{(2M_{2})^{2}}{27NU}\sum\limits_{\mathbf{k}}d^{'}(\mathbf{k})[i(1+\frac{36R_{2}^{2}}{M_{2}^{2}})\varphi_{6}^{'}(\mathbf{k})
+\frac{12R_{2}}{M_{2}}\varphi_{4}^{'}(\mathbf{k})].
\end{equation}

Two interesting features are clear in the above equations that determine $\Delta_1$ to $\Delta_6$. The first is that $\varphi_{i}^{'}(\mathbf{k})\neq\varphi_{i}(\mathbf{k})$ for $i$=2, 4, 6. From the above procedure of determining these pairing amplitudes we know that this is because the pairings with symmetry factors $\varphi_{2}(\mathbf{k})$, $\varphi_{4}(\mathbf{k})$ and $\varphi_{6}(\mathbf{k})$ are unequal amplitude superposition of $e^{i\mathbf{k}\cdot\boldsymbol{\delta}_{l}}$ ($l$=$\pm1$, $\pm2$, $\pm3$) factors. That is, the pairing fields on the $\pm\boldsymbol{\delta}_{3}$ bonds are stronger than those on the $\pm\boldsymbol{\delta}_{1}$ and $\pm\boldsymbol{\delta}_{2}$ bonds. The second feature is also discussed in the main text, that is the pairings $\Delta_{3}$ is explicitly mixed with $\Delta_{5}$ while $\Delta_{4}$ is explicitly mixed with $\Delta_{6}$. We have defined the concept of spatial-parity and orbital-parity in the main text. The above mixing of two kinds of pairings with opposite spatial-parity is a consequence of the interorbital superexchange term $H_{inter}$, which breaks explicitly the in-plane inversion symmetry of the correlation term. Another consequence of $H_{inter}$ is that, since the pairing potential is a time reversal invariant combination of even spatial-parity and odd spatial-parity components (see Eq.(5) of the main text), the self-consistent mean field superconducting solution consisting of several pairing components are time reversal invariant up to a global $U(1)$ phase. A time reversal invariant multi-component superconducting state was also found for iron pnictides.\cite{seo08}

\section{\label{sec:iGF} The iterative Green's function method}

Here, we explain how we get the surface Green's functions (GFs) in terms of the iterative GF method, which produce Figures 1 and 2 of the main text. First, we add a pairing term $\underline{\Delta}(\mathbf{k})$ to the normal state Hamiltonian $H_{0}(\mathbf{k})$. $\underline{\Delta}(\mathbf{k})$ can be one of the six spin singlet pairings defined in Sec.III (or, in the main text) or a specific linear combination of several pairing components. Introducing the Nambu basis $\psi^{\dagger}_{\mathbf{k}}=[\phi^{\dagger}_{\mathbf{k}},(\phi_{-\mathbf{k}})^{\textbf{T}}]$, we get the Bogoliubov-de Gennes (BdG) Hamiltonian as
\begin{equation}
H(\mathbf{k})=
\begin{pmatrix} H_{0}(\mathbf{k})-\mu I_{4} & \underline{\Delta}(\mathbf{k}) \\
-\underline{\Delta}^{\ast}(-\mathbf{k}) & \mu I_{4}-H^{\ast}_{0}(-\mathbf{k})
\end{pmatrix},
\end{equation}
where $\mu$ is the chemical potential and $I_{4}$ is the fourth-order unit matrix. The bulk GF is defined simply as $G_{b}(\mathbf{k},\omega)=[(\omega+i\eta)I_{8}-H(\mathbf{k})]^{-1}$, where $\eta$ is the positive infinitesimal and $I_{8}$ is the eighth-order unit matrix. In actual calculations, $\eta$ is taken as a small finite positive number (e.g., 10$^{-5}$ eV is used in this work).

To study the surface states living on the $xy$ surface, we consider a sample of the Bi$_{2}$X$_{3}$ (X is Se or Te) superconductor occupying the lower half space ($z<0$). The corresponding model is obtained by transforming the $z$ direction of the bulk model, Eq. (1) in the main text, from wave vector space to real space. Introducing an integral label $n$ to represent the various quintuple layers, with bigger $n$ indicating a larger $z$ coordinate, we can write the model as $\hat{H}=\hat{H}_{xy}+\hat{H}_{z}$, where $\hat{H}_{xy}$ contains the intra-quintuple-layer terms and $\hat{H}_{z}$ consists of the inter-quintuple-layer hopping terms. Denoting the Nambu basis in terms of the layer label $n$ and the two dimensional wave vectors $\tilde{\mathbf{k}}$ defined on the $k_{x}k_{y}$ plane, we have
\begin{eqnarray}
&&\hat{H}_{xy}=
\frac{1}{2}\sum\limits_{n\tilde{\mathbf{k}}}\psi^{\dagger}_{n\tilde{\mathbf{k}}}\begin{pmatrix} H^{'}_{0}(\tilde{\mathbf{k}})-\mu I_{4} & \underline{\Delta}(\tilde{\mathbf{k}}) \\
-\underline{\Delta}^{\ast}(-\tilde{\mathbf{k}}) & \mu I_{4}-H^{'\ast}_{0}(-\tilde{\mathbf{k}})
\end{pmatrix}\psi_{n\tilde{\mathbf{k}}}   \notag \\
&&=\frac{1}{2}\sum\limits_{n\tilde{\mathbf{k}}}\psi^{\dagger}_{n\tilde{\mathbf{k}}}h_{xy}(\tilde{\mathbf{k}})\psi_{n\tilde{\mathbf{k}}},
\end{eqnarray}
where
\begin{eqnarray}
H^{'}_{0}(\tilde{\mathbf{k}})&=&\epsilon^{'}(\tilde{\mathbf{k}})I_{4}+M^{'}(\tilde{\mathbf{k}})\Gamma_5
+A_{0}[c_{y}(\tilde{\mathbf{k}})\Gamma_{1}   \notag \\
&&-c_{x}(\tilde{\mathbf{k}})\Gamma_{2}]+R_{1}d_{1}(\tilde{\mathbf{k}})\Gamma_{3}+R_{2}d_{2}(\tilde{\mathbf{k}})\Gamma_{4}.
\end{eqnarray}
While the dependencies of $c_{x}$, $c_{y}$, $d_{1}$ and $d_{2}$ on the wave vectors keep unchanged, $\epsilon^{'}(\tilde{\mathbf{k}})=C_{0}+2C_{1}
+\frac{4}{3}C_{2}[3-\cos(\tilde{\mathbf{k}}\cdot\boldsymbol{\delta}_{1})-\cos(\tilde{\mathbf{k}}\cdot\boldsymbol{\delta}_{2})
-\cos(\tilde{\mathbf{k}}\cdot\boldsymbol{\delta}_{3})]$ and $M^{'}(\tilde{\mathbf{k}})=M_{0}+2M_{1}
+\frac{4}{3}M_{2}[3-\cos(\tilde{\mathbf{k}}\cdot\boldsymbol{\delta}_{1})-\cos(\tilde{\mathbf{k}}\cdot\boldsymbol{\delta}_{2})
-\cos(\tilde{\mathbf{k}}\cdot\boldsymbol{\delta}_{3})]$. The inter-quintuple-layer terms are
\begin{eqnarray}
&&\hat{H}_{z}=
\frac{1}{2}\sum\limits_{n\tilde{\mathbf{k}}}\psi^{\dagger}_{n\tilde{\mathbf{k}}}\begin{pmatrix} H_{0z} & 0 \\
0 & -H^{\ast}_{0z}
\end{pmatrix}\psi_{n+1,\tilde{\mathbf{k}}}+\text{H.c.}   \notag \\
&&=\frac{1}{2}\sum\limits_{n\tilde{\mathbf{k}}}\psi^{\dagger}_{n\tilde{\mathbf{k}}}h_{z}\psi_{n+1,\tilde{\mathbf{k}}}+\text{H.c.},
\end{eqnarray}
where $\text{H.c.}$ means taking the Hermite conjugation of the terms explicitly written out, and
\begin{equation}
H_{0z}=-M_{1}\Gamma_{5}-\frac{i}{2}B_{0}\Gamma_{4}.
\end{equation}

With the above model at hand, the surface GF for the surface layer of the semi-infinite sample occupying the $z<0$ half space is obtained iteratively by\cite{wang10,hao11}
\begin{equation}
G^{(m)}_{s}(\tilde{\mathbf{k}},\omega)=[g^{-1}-h^{\dagger}_{z}G^{(m-1)}_{s}h_{z}]^{-1},
\end{equation}
where $g=[(\omega+i\eta)I_{8}-h_{xy}(\tilde{\mathbf{k}})]^{-1}$ is the GF for an isolated quintuple layer. The superscripts $m$ and $m-1$ label the iteration steps. To begin, we set $G^{(0)}_{s}=g$ and get $G^{(1)}_{s}$. Then $G^{(1)}_{s}$ is put into the right side of Eq.(D6) to give $G^{(2)}_{s}$. The iteration is repeated until the difference between every corresponding component of $G^{(m)}_{s}$ and $G^{(m-1)}_{s}$ is smaller than a certain small positive number, which is set by hand to control the precision.

Once we have obtained the surface GF $G_{s}(\tilde{\mathbf{k}},\omega)$ in terms of the above iterative GF method (or, transfer matrix method), we can get the surface spectral function by summing up the imaginary parts of the four particle surface GFs as (note that, the wave vectors in the surface BZ are denoted as $\mathbf{k}_{xy}$ in the main text)
\begin{equation}
A(\tilde{\mathbf{k}},\omega)=-\frac{1}{\pi}\sum\limits_{i=1}^{4}\text{Im}G_{s,ii}(\tilde{\mathbf{k}},\omega).
\end{equation}

\end{appendix}


\end{document}